\documentclass[journal,12pt,onecolumn,draftclsnofoot,]{IEEEtran}
\usepackage{graphicx}
\usepackage{latexsym}
\usepackage{amsmath}
\usepackage{amsthm}
\usepackage{makecell,rotating,multirow,diagbox}
\usepackage{amsfonts}
\usepackage{subfigure}
\usepackage{dsfont}
\usepackage{epstopdf}
\usepackage{threeparttable}
\usepackage{multirow}
\usepackage{enumerate}
\usepackage{epsfig}
\usepackage{bm}
\usepackage{cite}
\usepackage{algorithmic}
\usepackage{textcomp}
\usepackage{xcolor}
\usepackage[linesnumbered, ruled]{algorithm2e}
\usepackage{amssymb}


\usepackage{enumitem}

\ifCLASSINFOpdf
\else
\fi
\begin{document}
%
\title{Joint Model and Data Driven Receiver Design for Data-Dependent Superimposed Training Scheme with Imperfect Hardware}

%
%
%

\author{Chaojin~Qing,~\IEEEmembership{Member,~IEEE,}
        Lei~Dong,
        Li~Wang,
        Jiafan~Wang,
        and~Chuan~Huang,~\IEEEmembership{Member,~IEEE}

\thanks{C. Qing, L. Dong and L. Wang are with the School of Electrical Engineering and Electronic Information, Xihua University, Chengdu, 610039, China (E-mail: qingchj@mail.xhu.edu.cn). }
\thanks{J. Wang is with the Synopsys Inc., 2025 NE Cornelius Pass Rd, Hillsboro, OR 97124, USA (E-mail: jifanw@gmail.com).}
\thanks{C. Huang is with the School of Science and Engineering, The Chinese University of Hong Kong, Shenzhen, 518172, China (E-mail: huangchuan@cuhk.edu.cn).}}

\maketitle

\begin{abstract}
Data-dependent superimposed training (DDST) scheme has shown the potential to achieve high bandwidth efficiency, while encounters symbol misidentification caused by hardware imperfection. To tackle these challenges, a joint model and data driven receiver scheme is proposed in this paper. Specifically, based on the conventional linear receiver model, the least squares (LS) estimation and zero forcing (ZF) equalization are first employed to extract the initial features for channel estimation and data detection. Then, shallow neural networks, named CE-Net and SD-Net, are developed to refine the channel estimation and data detection, where the imperfect hardware is modeled as a nonlinear function and data is utilized to train these neural networks to approximate it. Simulation results show that compared with the conventional minimum mean square error (MMSE) equalization scheme, the proposed one effectively suppresses the symbol misidentification and achieves similar or better bit error rate (BER) performance without the second-order statistics about the channel and noise.

\end{abstract}


%
\IEEEpeerreviewmaketitle
\section{Introduction}
\label{sec:introduction}
\IEEEPARstart{D}{ata}-dependent superimposed training (DDST) scheme has triggered wide research interests due to its high bandwidth efficiency \cite{Ref_4,rr8,rr10,rr9,rr6,RefZZ_1,RefZZ_2,rr5,Ref_1}. Relative to the conventional superimposed training (ST) scheme \cite{I-a3}, the training sequence in DDST scheme is combined with a judiciously selected data-dependent sequence. With this modified training sequence, the DDST scheme can efficiently eliminate the data-induced interference and achieve better performance for channel estimation and data detection \cite{Ref_4,rr8,rr10,rr9,rr6,RefZZ_1,RefZZ_2,rr5,Ref_1}. In essence, the DDST scheme fully cancels the effects of the unknown transmission data on the performance of the channel estimator, and this leads to various methods proposed to improve the data detection performance \cite{r4}. However, some information-bearing data at certain frequency bins are removed prior to transmission, preventing linear receiver recovery of transmitted symbols and resulting in symbol misidentification \cite{rr8}. Meanwhile, wireless communication systems with imperfect devices or blocks, e.g., high power amplifier (HPA), and digital-to-analog converter (DAC), will inevitably cause nonlinear distortion to the received signals \cite{Ref_3}, and give rise to great challenges for existing schemes (e.g., the DDST scheme in [9]). For a mobile device, the nonlinear distortion may be more serious due to its low-cost components \cite{Ref722_1}, limited computational resources \cite{Ref722_3}, and low complexity processing methods \cite{Ref722_4}, etc. The nonlinear distortion destroys the orthogonality of the training sequence \cite{Conf_Zjiao} and confuses the modulation constellation \cite{Ref2_Conf25}, so that the performance of the channel estimation and data detection degrades seriously at the receiver side, which further aggravates symbol misidentification.

To guarantee the benefits of DDST schemes, nonlinear distortion has to be considered during practical applications. The previous study \cite{Ref_3} suggests that machine learning, in particular deep learning (DL), is prominent to solve nonlinear problems. In recent years, DL-based physical-layer technique shows its promising prospects in wireless communication system, e.g., signal detection \cite{c9},\cite{TWC_1}, precoding \cite{c7}, channel state information (CSI) feedback \cite{c10}, \cite{TWC_2}, and channel estimation \cite{c11}, \cite{c4}, etc. For channel estimation and data detection, however, there are limited works focused on the DL-based DDST scheme and even less on the nonlinear distortion caused by hardware imperfection. Since the nonlinear distortion inevitably appears in wireless communication systems \cite{a8}, to guarantee the benefit of high bandwidth efficiency of the DDST scheme, we investigate a joint model and data driven receiver with the consideration of nonlinear distortion in this paper.


\subsection{Related Works}

Relative to the ST scheme, the DDST scheme improved the overall performance of channel estimation and data detection \cite{Ref_4}. But, the partial information of transmitted data symbols is discarded in the frequency domain, and thus the data detection is degraded due to the symbol misidentification. In other words, the channel estimation accuracy of DDST scheme is obtained at the cost of degradation of data detection performance. To relieve this problem, various methods have been proposed in \cite{rr8,rr10,rr9,rr6,RefZZ_1,RefZZ_2,rr5,Ref_1}.

In \cite{rr8}, authors raised a partially DDST (PDDST) scheme to reduce symbol misidentification by investigating the trade-off between interference cancellation and frequency integrity. In reality, to improve the data detection performance, partial performance of channel estimation was sacrificed in the scheme of PDDST. This scheme was applied to the systems of orthogonal frequency-division multiplexing in \cite{ rr10} with position shifting of pilot signals and the restriction of the lowest power sum. In \cite{rr9}, a semi-blind approach was proposed for channel estimation and data detection, in which the first-order statistics of received data were developed for an iteration approach with kernel weighted least squares (LS). With the increased computational complexity, the semi-blind approach proposed in \cite{rr9} enhances the accuracy of channel estimation and therefore improves the data detection performance. By considering the equi-spaced (square quadrature) amplitude modulation, \cite{rr6} investigated the data coding method to remove the error floor of data detection in DDST scheme. In \cite{rr6}, the discarded symbol information was retrieved according to the relevancy between encoding and decoding, which avoids partial symbol misidentification. In \cite{RefZZ_1}, to avoid symbol misidentification, a constellation rotation method was proposed and used to preserve the partial symbol information discarded by DDST scheme. By using the signal sub-space technique, \cite{RefZZ_2} theoretically analyzed the phenomenon of symbol misidentification (also called data identification problem). It was shown mathematically that the occurrence of symbol misidentification was related to both the modulation scheme and the pilot pattern. The precoding-based approach is another family to improve data detection \cite{rr5} \cite{Ref_1}. By developing the inter-symbols correlation, the precoding-based approach can effectively retrieve the discarded symbol information. In \cite{rr5}, the Zadoff-Chu sequence-based precoding matrix was exploited to resolve the symbol misidentification. In particular, the detailed design criteria for the precoding matrix were presented in \cite{Ref_1}, e.g., the precoding matrix should have the form of the Kronecker product of an identity matrix and a diagonal matrix with equal-magnitude diagonal elements, which presents good guidance.

Relative to the DDST scheme, except for the semi-blind approach in \cite{rr9}, the training sequence-based improvements mentioned above require the transmitter to be modified. In fact, the transmitter of the mobile device faces the dilemmas of limited battery energy, storage space and computational resources. This is impractical to modify the transmitter of mobile device by increasing the processing complexity and storage space. Further, the nonlinear distortion caused by imperfect hardware is usually time-varying, which makes the transmitter modification ineffective since the receiver has to deal with the nonlinear distortion again. In this paper, we investigate a receiver scheme (i.e., without the modification of transmitter) for base station (BS) or access point (AP) to improve the symbol misidentification of DDST with the consideration of nonlinear distortion, which is different from the existing schemes. To the best of our knowledge, very limited works discuss the impact of nonlinear distortion on the DDST scheme, the receiver scheme to improve the DDST, and the application of DL technology to reduce symbol misidentification. Thus, we investigate a joint model and data driven receiver scheme to improve the performance of DDST scheme affected by nonlinear distortion.

\subsection{Contributions}

Focusing on the nonlinear distortion scenarios, the joint model and data driven receiver scheme is proposed to improve channel estimation and data detection in DDST. Our key insight is to view the hardware imperfection and unsatisfactory recovery of the linear receiver in DDST as a nonlinear problem. Then, the problem is handled by DL technology which shows its prominent ability in dealing with nonlinearity. We combine the advantages of linear receiver and nonlinear solution for tackling the symbol misidentification in DDST scheme. Specifically, based on the conventional model driven modes, the LS estimation and zero forcing (ZF) equalization are first employed to highlight the initial linear features for channel estimation and data detection. Then, with obtained initial features, we construct the shallow neural networks, named as CE-Net and SD-Net, to refine the channel estimation and data detection. Here, continuing with the linear receiver solutions, the data driven mode is exploited to refine the channel estimation accuracy and data detection performance by solving a nonlinear problem, thereby improving the symbol misidentification. Our investigation remedies the deficiencies of the existing DDST-based scheme which is facing difficulties in application due to the lack of considerations of hardware imperfection.


The remainder of this paper is structured as follows: In Section II, we present the system model of DDST scheme with nonlinear distortion. The proposed receiver scheme is presented in Section III, and the numerical results are given in Section IV. Finally, Section V concludes our work.

\textit{Notations}: Bold face upper case and lower case letters denote matrix and vector, respectively. ${\left(\cdot \right)^T}$, ${\left(\cdot \right)^H}$, $\mathbf{(\cdot)^\dag}$, stand for the transpose, conjugate transpose, and matrix pseudo-inverse, respectively. $\bigotimes$ is the Kronecker product. ${\mathop{\mathrm{Re}(\cdot)}}$ and ${\mathop{\mathrm{Im}(\cdot)}}$ denote the real and imaginary parts of a complex number, respectively. $\mathbf{I}_N$ represents an $N \times N$ identity matrix. ${\mathrm{diag}(\cdot)}$ is the diagonalization operation of matrix. ${\mathrm {{\mathbb{BN}}}}\left( \cdot \right)$ denotes the operation of batch normalization. ${\left\|  \cdot  \right\|_2}$ is the Euclidean norm. ${\mathbf{F}_N}$ is the normalized $N \times N$ fourier transform matrix \cite{Ref_1}. $\max \left( {a,b} \right)$ stands for taking the larger of $a$ and $b$. $\mathbf{x}(M:N)$ represents the sub-vector of $\mathbf{x}$ by extracting $\mathbf{x}$ from its $M$th entry to $N$th entry.


\section{System Model}
\begin{figure*}[t]
\centering
\includegraphics[scale=0.83]{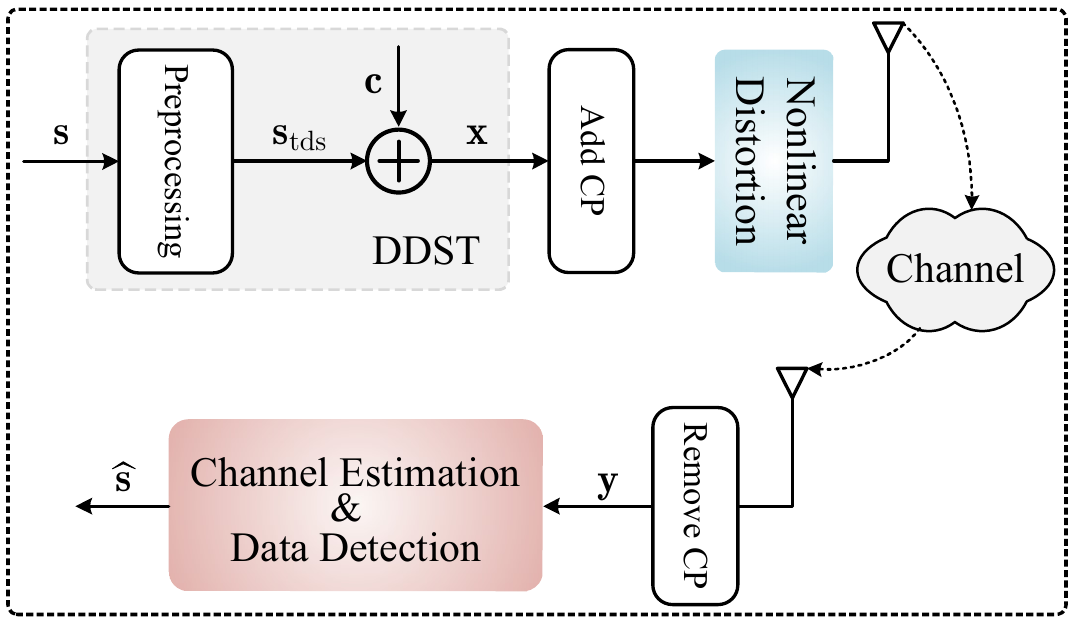}
\caption{System model for the DDST scheme with imperfect hardware.}
\label{fig1}
\end{figure*}

As shown in Fig.~\ref{fig1}, this paper considers a cyclic prefix (CP)-protected single carrier system with the DDST scheme. At the transmitter, after the DDST processing and adding CP, the transmitted signal experiences nonlinear distortion caused by imperfect hardware. At the receiver, after the CP is removed, the channel estimation and data detection are then performed. In the following, we first describe the signal model at the transmitter, and then present the received signal model.


\subsection{Transmit Signal Model}
\begin{figure*}[t]
\centering
\includegraphics[scale=0.85]{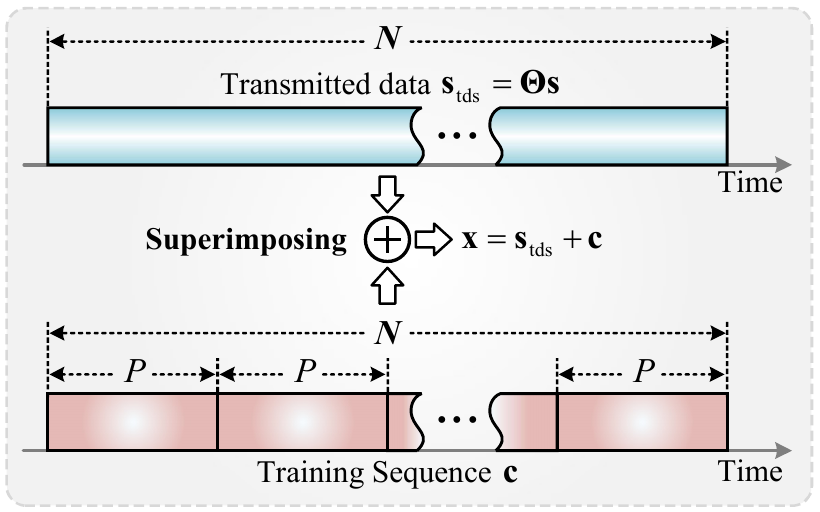}
\caption{Construction of one data frame for DDST.}
\label{figFrame}
\end{figure*}

Fig.~\ref{figFrame} shows how to construct one data frame with length $N$ at the transmitter. In the preprocessing phase, a frame of data $\mathbf{s} \in \mathbb{C}^{N \times 1}$ is first processed to obtain ${\mathbf{{s}}_\mathrm{tds}}$ \cite{Ref_1}, i.e.,
\begin{equation}\label{EQ4}
{{\mathbf{{s}}_\mathrm{tds}} = {\mathbf{\Theta s}}},
\end{equation}
where $\mathbf{\Theta}\in \mathbb{C}^{N\times N}$ is a DDST preprocessing matrix and is usually chosen as \cite{Ref_4}
\begin{equation}\label{EQ_theta}
\mathbf{\Theta} = {{\mathbf{I}_N} - {\mathbf{J}}}.
\end{equation}
Here, $\mathbf{J} \in \mathbb{C}^{N \times N}$ is subtracted from $\mathbf{I}_N$ to eliminate the interference of data sequence to training sequence, and is usually defined as \cite{Ref_4}
\begin{equation}\label{EQ7}
{{\mathbf{J}} =  \frac{1}{Q} {{\mathbf{J}}_Q} \otimes {{\mathbf{I}}_P},}
\end{equation}
where ${\mathbf{J}_Q}\in \mathbb{C}^{Q\times Q}$ is given as
\begin{equation}\label{EQ8}
{{{\mathbf{J}}_Q} = \left[ {\begin{array}{*{20}{c}}
1&{{e^{\frac{{j2\pi t\left( {Q - 1} \right)}}{Q}}}}&{{e^{\frac{{j2\pi t\left( {Q - 2} \right)}}{Q}}}}& \ldots &{{e^{\frac{{j2\pi t}}{Q}}}}\\
{{e^{\frac{{j2\pi t}}{Q}}}}&1&{{e^{\frac{{j2\pi t\left( {Q - 1} \right)}}{Q}}}}& \ldots &{{e^{\frac{{j2\pi 2t}}{Q}}}}\\
{{e^{\frac{{j2\pi 2t}}{Q}}}}&{{e^{\frac{{j2\pi t}}{Q}}}}&1& \ldots &{{e^{\frac{{j2\pi 3t}}{Q}}}}\\
 \vdots & \vdots & \vdots & \ddots & \vdots \\
{{e^{\frac{{j2\pi t\left( {Q - 1} \right)}}{Q}}}}&{{e^{\frac{{j2\pi t\left( {Q - 2} \right)}}{Q}}}}&{{e^{\frac{{j2\pi t\left( {Q - 3} \right)}}{Q}}}}& \ldots &1
\end{array}} \right]},
\end{equation}
with $Q$ and $P$ being constants, satisfying ${Q = \lfloor {N}/{P}\rfloor}$, and $t$ being the shift variable for constellation rotation \cite{RefZZ_1}.

To estimate the channel, a training sequence with length $N$, denoted as $\mathbf{c}$, is first constructed and then superimposed onto the processed data $\mathbf{s}_\mathrm{tds}$ to form the transmitted signal. Similar to \cite{Ref_4}, \cite{RefZZ_1} and \cite{Ref_1}, the training sequence $\mathbf{c}$ is constructed by $Q$ repetitive $P$-period random sequences. By superimposing ${\mathbf{{s}_\mathrm{tds}}}$ with ${\mathbf{c}}$, the transmitted signal $\mathbf{x} \in \mathbb{C}^{N \times 1}$ is then given as
\begin{equation}\label{EQ3}
{{\mathbf{x}} = {\mathbf{{s}_\mathrm{tds}}} + {\mathbf{c}}}.
\end{equation}
Considering the nonlinear distortion caused by imperfect hardware, the real transmitted signal is given as

\begin{equation}\label{EQ2}
{{\mathbf{x}_\mathrm{dis}} = {f_{\mathrm{dis}}}\left( {\mathbf{x}} \right)},
\end{equation}
where ${\mathbf{x}_\mathrm{dis}}$ denotes the distorted version of $\mathbf{x}$, and ${f_{\mathrm{dis}}}(\cdot)$ represents the nonlinear-distortion function \cite{Ref4_24}.

\subsection{Received Signal Model}
After the CP is removed, the received signal ${\mathbf{y}}$ with length $N$ is given as
\begin{equation}\label{EQ1}
{{\mathbf{y}} = {{\mathbf{H}_c}}{\mathbf{{x}_\mathrm{dis}}} + {\mathbf{v}},}
\end{equation}
where ${\mathbf{v}} \in \mathbb{C}^{N \times 1}$ denotes the circularly symmetric complex Gaussian (CSCG) noise with mean zero and variance ${\sigma _v^2}$, and ${\mathbf{H}_c}\in{{\mathbb{C}}^{N\times N}}$ is a cyclic matrix reflecting the multi-path channel fading effect \cite{Ref_1}. The first column of ${\mathbf{H}_c}$ is given as ${{\mathbf{h}} = {\left[ {{h_0}{\rm{, }} \cdots {\rm{ ,}}{h_{L - 1}},0, \ldots ,0} \right]^{\rm{T}}}}$
${\in {{\mathbb{C}}^{N\times 1}}}$, with ${L}$ being the number of channel paths. Then, the received signal $\mathbf{y}$ after removing CP is rewritten as
\begin{equation}\label{EQ10}
{{\mathbf{y}} = {{\mathbf{H}_c}}{f_{\mathrm{dis}}}\left( {\left( {{\mathbf{I}_N} - {\mathbf{J}}} \right){\mathbf{s}} + {\mathbf{c}}} \right) + {\mathbf{v}}.}
\end{equation}
\section{Channel Estimation and Data Detection}
In practical wireless communication systems, imperfect hardware is usually employed by mobile devices to reduce the hardware cost, while causing serious nonlinear distortion for transmitted signal \cite{Impef_Ref1}, \cite{Impef_Ref2}. Nonlinear distortion destroys the orthogonality between the received and local training sequence \cite{Conf_Zjiao} and confuses the modulation constellation \cite{Ref2_Conf25}, which seriously degrade the performance of channel estimation and data detection.

\begin{figure*}[t]
\centering
\includegraphics[scale=0.9]{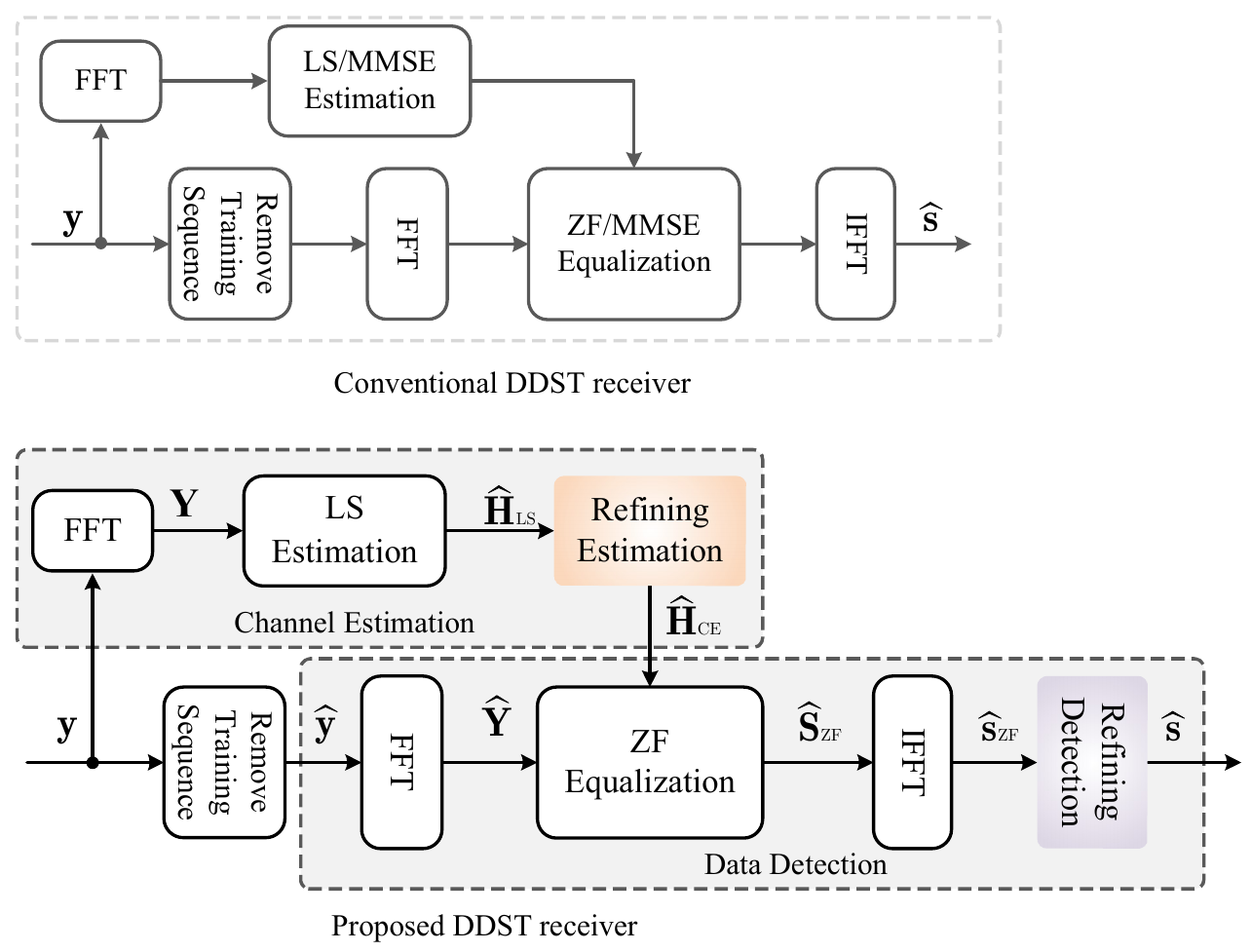}
\caption{Architecture of various DDST receivers.}
\label{fig_receiAch}
\end{figure*}

To tackle this issue, we propose a joint model and data driven scheme to design the receiver, which is shown in Fig.~\ref{fig_receiAch}. Based on the conventional structure of the receiver, the low-complexity and high-practicability LS estimation and ZF equalization are first employed to capture the initial features from the model driven channel estimation and data detection. Then, we embed simple and effective shallow networks to respectively refine the channel estimation and data detection by using the data driven method \cite{Ref722_5}. Different from the conventional receiver, the proposed receiver incorporates linear and nonlinear processing methods together by exploiting both the model and data driven approaches. Compared to the conventional receiver using  minimum mean square error (MMSE) channel estimation and MMSE equalization, our receiver can achieve similar or better detection performance. Besides, the second-order statistics about the channel and noise could be avoid, which are difficult to be obtained in practical applications \cite{Ref5_15}.

%


In the following subsections, we first elaborate the channel estimation in Section III-{A}. Then, in Section III-{B}, the details of data detection are described. Finally, the online deployment is shown in Section III-{C}.

\subsection{Channel Estimation}
\begin{figure*}[t]
\centering
\includegraphics[scale=0.9]{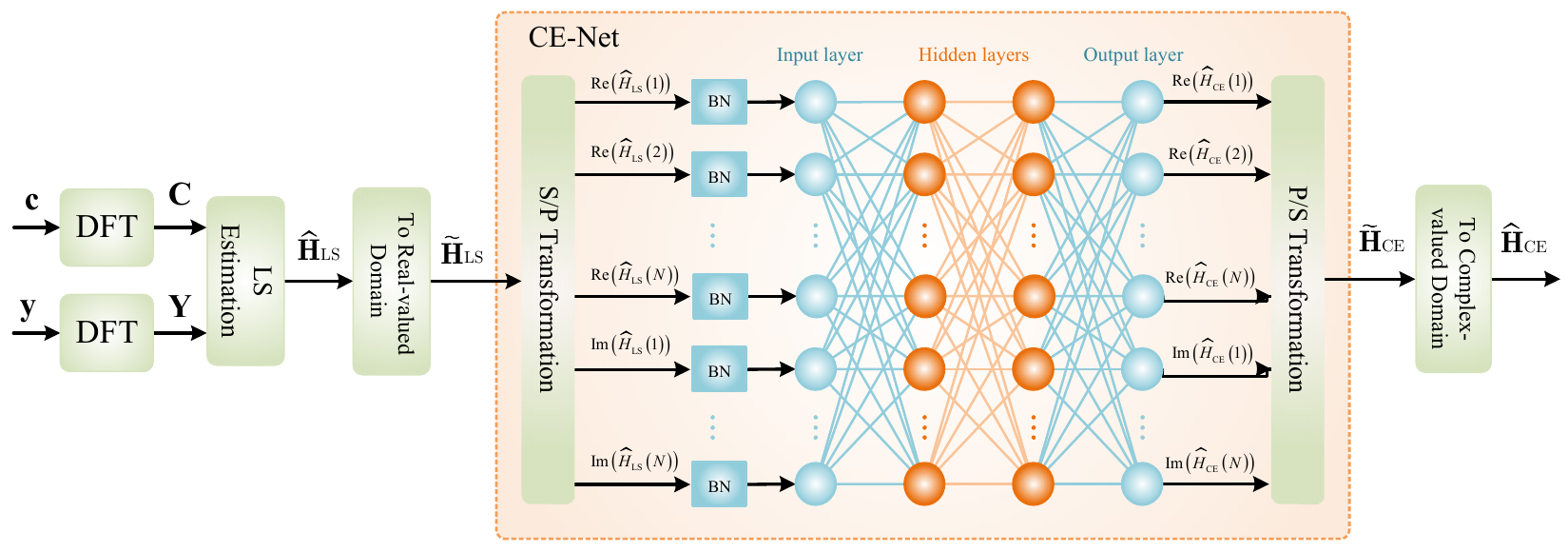}
\caption{Channel estimation for DDST receiver.}
\label{fig2}
\end{figure*}

The channel estimation for the proposed DDST receiver scheme is shown in Fig.~\ref{fig2}, in which the LS estimation is first performed and then followed by a shallow network named as CE-Net to refine the channel estimation.

\subsubsection{LS Estimation}
We employ LS estimation to obtain an initial channel estimation in the frequency domain, which also serves as the input of the CE-Net. The received signal ${\mathbf{y}}$ given in (\ref{EQ10}) and the training sequence ${\mathbf{c}}$ are first transformed into the frequency domain, i.e.,
\begin{equation}\label{EQ13_add}
 {\begin{array}{*{20}{c}}
{{\mathbf{Y}} = {{\mathbf{F}}_N}{\mathbf{y}}},\\
{{\mathbf{C}} = {{\mathbf{F}}_N}{\mathbf{c}}},
\end{array}}
\end{equation}
where $\mathbf{Y}$ and $\mathbf{C}$ are the frequency domain representations of received signal $\mathbf{y}$ and training sequence $\mathbf{c}$, respectively. According to  the LS algorithm \cite{Ref_4}, the frequency domain channel estimator ${{\mathbf{\widehat{H}}_P}} \in \mathbb{C} ^ {P \times 1} $ is given as
\begin{equation}\label{EQ15_add}
{{\widehat{\mathbf{H}}}_{P}} = {\left[ {\frac{{Y\left( 0 \right)}}{{C\left( 0 \right)}},\frac{{Y\left( Q \right)}}{{C\left( Q\right)}}, \ldots, \frac{{Y\left( {\left( {P - 1} \right)Q} \right)}}{{C\left( {\left( {P - 1} \right)Q} \right)}}} \right]^T},
\end{equation}
where $Y(k)$ and $C(k)$, $k= 0, \ldots, N-1$, are frequency domain values for the received signal $\mathbf{y}$ and the training sequence $\mathbf{c}$, respectively. Then, we transform ${{\mathbf{\widehat{H}}_P}}$ into the time domain, and obtain the time domain channel estimator ${\widehat{\mathbf{h}}} \in \mathbb{C} ^ {P \times 1}$ as
\begin{equation}\label{EQ15_add111}
{{{{\widehat{\mathbf{h}}}}} = {\mathbf{F}}_P^{H}{{\mathbf{\widehat{H}}}_P}},
\end{equation}
and add zero at the end of ${\widehat{\mathbf{h}}}$ to form a vector $\mathbf{\widetilde{h}}$ with length $N$, i.e.,
\begin{equation}\label{EQ15_add222}
{{\mathbf{\widetilde{h}}} = {\left[ {{\mathbf{{\widehat{\mathbf{h}}}}}^T,\underbrace {0, \ldots ,0}_{N - P}} \right]^T}}.
\end{equation}
The LS estimation $\mathbf{\widehat{H}}_{\mathrm{LS}}$ is obtained by transforming $\mathbf{\widetilde{h}}$ into the frequency domain, i.e.,
\begin{equation}\label{EQ15_add333}
{{{\mathbf{\widehat{H}}}_{\mathrm{LS}}} = {{\mathbf{F}}_N}{\mathbf{\widetilde{h}}}}.
\end{equation}

%
%
%
%
%


\subsubsection{Refining Estimation}
After the LS estimation, we adopt the CE-Net to refine the channel estimation by alleviating the influence of nonlinear distortion. The main steps are explained as follows.

\textbf{Constructing CE-Net:} To construct a neural network, the network layers and the neurons of each layer usually need to be first determined \cite{RRef1111}. In CE-Net, the neurons and layers are first set by referencing the methods in \cite{Ref2424}, and then perform a fine parameter tuning according to a large number of experiments. That is, the architecture of CE-Net is constructed to capture a trade-off between its structure complexity and the system's bit error rate (BER) performance.
The CE-Net is presented in Fig.~\ref{fig2}, which consists of an input layer, two hidden layers, and an output layer. Table~\ref{tableQ_1} summarizes the network architecture of CE-Net, and its detailed descriptions are given as follows.

\begin{table}[t]
\renewcommand{\arraystretch}{1}
\caption{Architecture of CE-Net}
\label{tableQ_1} \centering
\begin{tabular}{c|c|c|c|c}
\Xhline{1.3pt}
Layer & Input   & Hidden 1 & Hidden 2 & Output \\
\Xhline{1.3pt}
Number of neurons   & ${2N}$  & ${2N}$  & ${2N}$  & ${2N}$    \\
\hline
Batch normalization & $\surd$         & $\times$    & $\times$   & $\times$ \\
\hline
Activation function & None     & ReLU    &ReLU    & Linear \\
\hline

\end{tabular}
\end{table}

\begin{itemize}
  \item
  The neurons of the input layer, hidden layers, and output layer are all set as ${2N}$ to simplify the network structure and accelerate the network training.
\end{itemize}

\begin{itemize}
  \item
  In order to accelerate the convergence of the network training and prevent the overfitting problem \cite{a9}, batch normalization (BN) is employed to normalize the network inputs.
\end{itemize}

\begin{itemize}
  \item
 The activation function rectified linear unit (ReLU) defined as ${f_a}\left( x \right) = \max \left( {0,x} \right)$, is adopted for the two hidden layers due to its superiority in alleviating the gradient vanishing problem \cite{a10}. For the output layer, the linear activation function is employed \cite{c10}.
\end{itemize}


\textbf{Channel estimation:} We refine the channel estimation by exploiting the CE-Net, which is given in Fig. \ref{fig2}. Usually, signals in common wireless systems are complex-valued, and this brings difficulty to the DL-based network framework which requires a real-valued data set. To address this issue, the complex-valued ${{\mathbf{\widehat{H}}}_{\mathrm{LS}}} \in{\mathbb{C}}^{N\times1}$ is reshaped to real-valued ${{\mathbf{\widetilde{H}}}_{\mathrm{LS}}} \in{\mathbb{R}}^{2N\times1}$, i.e.,
 \begin{equation}\label{EQ16_add111}
{{{\bf{\widetilde{H}}}_{\mathrm{LS}}} = {{\left[ {{\mathop{\rm Re}\nolimits} \left( {{\bf{\widehat{H}}}_{\mathrm{LS}}^T} \right),{\mathop{\rm Im}\nolimits} \left( {{\bf{\widehat{H}}}_{\mathrm{LS}}^T} \right)} \right]}^T}}.
 \end{equation}
Then, the entries of ${{\bf{\widetilde{H}}}_{\mathrm{LS}}}$, i.e., $\mathop{\rm Re} \left( \widehat{H}_{\mathrm{LS}}(1) \right)$, $\mathop{\rm Re} \left( \widehat{H}_{\mathrm{LS}}(2) \right)$, $\cdots$, $\mathop{\rm Re} \left( \widehat{H}_{\mathrm{LS}}(N) \right)$, $\mathop{\rm Im} \left( \widehat{H}_{\mathrm{LS}}(1) \right)$, $\cdots$, $\mathop{\rm Im} \left( \widehat{H}_{\mathrm{LS}}(N) \right)$, compose the inputs of CE-Net. By using the CE-Net, the refined CSI, denoted as ${{\mathbf{\widetilde{H}}}_{\mathrm{CE}}} \in{\mathbb{R}}^{2N\times1}$, is formulated as
\begin{equation}\label{EQ16_add}
{\mathbf{\widetilde{H}}_{\mathrm{CE}}} = {f^{\left( {\cal L} \right)}}\left( {\mathbf{W}_{\mathrm{CE}}^{\left( {\cal L} \right)}{f^{\left( {{\cal L} - 1} \right)}}\left( { \cdots {f^{\left( 2 \right)}}\left( {\mathbf{W}_{\mathrm{CE}}^{\left( 2 \right)}{\mathbf{\widetilde{H}}_{\mathrm{LS}}} + \mathbf{b}_{\mathrm{CE}}^{\left( 2 \right)}} \right)} \right) + \mathbf{b}_{\mathrm{CE}}^{\left( {\cal L} \right)}} \right),
\end{equation}
where ${f^{\left( \ell  \right)}}$, $\mathbf{W}_{\mathrm{CE}}^{\left( \ell  \right)}$, and $\mathbf{b}_{\mathrm{CE}}^{\left( \ell  \right)}$ are the activation function, weight matrix and bias vector of the $\ell$th layer $\forall \ell  = 2, \cdots ,{\cal L}$, respectively. According to (\ref{EQ16_add}), we solve the nonlinear problem to suppress the nonlinear distortion and obtain the refined CSI without using the second-order statistics about channel and noise.

\textbf{Model Training:} After constructing the CE-Net architecture and determining the network output, we need to train the CE-Net. The training details, i.e., training data collection, loss function description, and parameter initialization, are described as follows.
\begin{itemize}
  \item \textit{Data collection:} For the training of the CE-Net, the training set $\left\{ {{{\mathbf{\widetilde{H}}}_{\mathrm{LS},}}{\mathbf{\widetilde{H}}_{\mathrm{Label}}}} \right\}$ is generated as follows.

      \hspace{1.3em}Considering the frequency-selective fading channel \cite{Ref_1}, $\mathbf{h}$ defined in (\ref{EQ1}) is generated according to the widely adopted channel model COST2100 \cite{a7_add} without loss of generality, where the outdoor semi-urban scenario at the 300MHz band is considered. Then, $\mathbf{h}$ is transformed into the frequency domain to form label ${\mathbf{{H}}_{\mathrm{Label}}}$, and we map the complex-valued ${\mathbf{{H}}_{\mathrm{Label}}}$ to real-valued ${\mathbf{\widetilde{H}}_{\mathrm{Label}}}$.  After we obtain the receiver signal $\mathbf{y}$ from (\ref{EQ10}), the input of the CE-Net ${{\mathbf{\widehat{H}}}_{\mathrm{LS}}}$ is generated according to (\ref{EQ13_add})--(\ref{EQ15_add333}). Finally, the complex-valued set of $\{ {\mathbf{\widehat{H}}}_{\mathrm{LS}} \}$ is reshaped to the real-valued set $\{ {\mathbf{\widetilde{H}}}_{\mathrm{LS}} \}$. We use training set $\left\{ {{{\mathbf{\widetilde{H}}}_{\mathrm{LS},}}{\mathbf{\widetilde{H}_{\mathrm{Label}}}}} \right\}$ to train CE-Net. In addition, to validate the trained network parameters during the training phase, a validation set is also generated by using the same generation method of training set, and thus we can capture a set of optimized network parameters.

\end{itemize}

\begin{itemize}
\item \textit{Loss function:} To train CE-Net, mean squared error (MSE) is employed as the loss function, which is defined as
 \begin{equation}\label{EQ18}
{{{L}}_{{\mathrm{CE - Net}}} = \frac{1}{{{T_1}}}\left\| {\mathbf{\widetilde{H}}_{\mathrm{Label}} - {\mathbf{\widetilde{H}}_{\mathrm{CE}}}} \right\|_2^2 + {\alpha _{\mathrm{CE}}}\sum\limits_{\ell = 2}^4 {{{\left\| {\mathbf{W}_{\mathrm{CE}}^{\left( \ell  \right)}} \right\|}_2^2}} },
\end{equation}
where ${T_1}$ denotes the number of training samples, and $\alpha _{\mathrm{CE}}$ represents the regularization coefficient. Usually, the regularization item is used to prevent overfitting and improve the generalization performance \cite{DeppRef}. Moreover, CE-Net's performance (e.g., the CE-Net's estimation accuracy) is affected by $\alpha _{\mathrm{CE}}$. Thus, the regularization coefficient $\alpha _{\mathrm{CE}}$ needs to be optimized, which will be discussed in Section IV-{B}.
\end{itemize}

\begin{itemize}
  \item \textit{Weight and bias initialization:} From \cite{a12}, appropriate initialization can effectively avoid the gradient exploding or vanishing problem. For expression convenience, we use $\left\{ {{{\mathbf{W}}_{\mathrm{CE}}^{\left( \ell  \right)}}; {{\mathbf{b}}_{\mathrm{CE}}^{\left( \ell  \right)}}} \right\}$, ${\ell} = 2,3,4$, to represent the set of training parameters. For CE-Net training, elements of $\left\{ {{{\mathbf{W}}_{\mathrm{CE}}^{\left( \ell  \right)}}; {{\mathbf{b}}_{\mathrm{CE}}^{\left( \ell  \right)}}} \right\}$ are initialized as independent and identically distributed (i.i.d) Glorot uniform distribution \cite{a13}.

\end{itemize}

\subsection{Data Detection}
\begin{figure*}[t]
\centering
\includegraphics[scale=0.75]{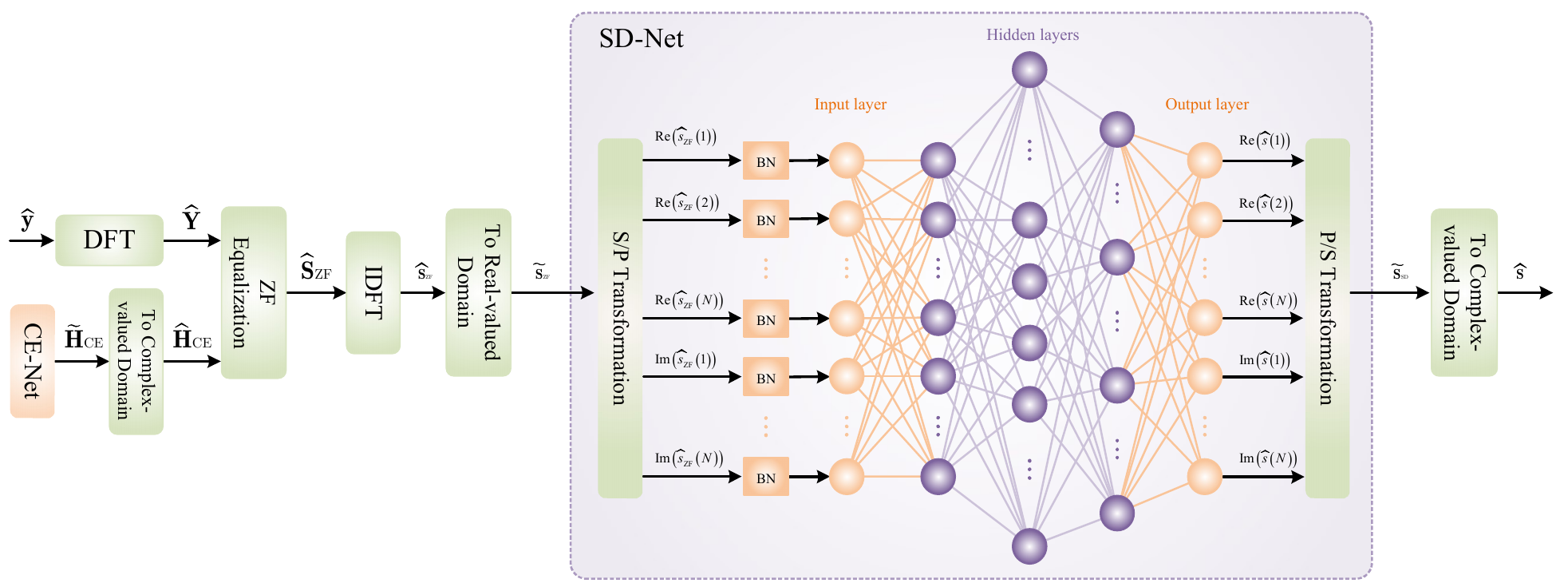}
\caption{Data detection for DDST receiver.}
\label{fig222}
\end{figure*}

The data detection for the proposed DDST receiver scheme is given in Fig.~\ref{fig222}. To avoid using the second-order statistics of the noise, the ZF equalization is first employed, followed by a shallow network named as SD-Net to refine the detection performance of DDST scheme.

\subsubsection{ZF Equalization}
From (\ref{EQ3}), the training sequence $\mathbf{c}$ is superimposed onto the transmitted data sequence $\mathbf{s}$. To recover the data sequence, the training sequence $\mathbf{c}$ has to be removed. Based on the DDST scheme \cite{Ref_1}, we remove the training sequence $\mathbf{c}$ according to

\begin{equation}\label{EQ_EQ3333}
{{\mathbf{\widehat{y}}} = \left( {{\mathbf{I}_N} - {\mathbf{J}}} \right){\mathbf{y}}},
\end{equation}
where $\mathbf{\widehat{y}} \in{\mathbb{C}}^{N\times1}$ denotes the recovery of the received signal $\mathbf{y}$. In this paper, the frequency domain equalization is utilized to combat the channel effect \cite{Ref_1}. Thus, ${\mathbf{\widehat{y}}}$ in (\ref{EQ_EQ3333}) is first transformed to the frequency domain to form $\mathbf{\widehat{Y}} \in{\mathbb{C}}^{N\times1}$ as
\begin{equation}\label{EQ_EQ4444}
{{\mathbf{\widehat{Y}}} = {{\mathbf{F}}_N}{\mathbf{\widehat{y}}}}.
\end{equation}
With the obtained ${\mathbf{\widehat{Y}}}$, the ZF equalization \cite{Ref2424} is employed to highlight the initial features of linear receiver for data detection. Based on the ${{\mathbf{\widehat{H}}}_{\mathrm{CE}}}$ (i.e., the refined CSI of CE-Net) and $\mathbf{\widehat{Y}}$, the ZF-based equalization is given by
\begin{equation}\label{EQ_EQ555}
{{\mathbf{\widehat{S}}_{\mathrm{ZF}}} = {{\mathbf{G}}_\mathrm{ZF}}{\mathbf{\widehat{Y}}}},
\end{equation}
where ${\mathbf{\widehat{S}}_{\mathrm{ZF}}}$ denotes the output of the ZF equalization, ${{\mathbf{G}}_\mathrm{ZF}} \in{\mathbb{C}}^{N \times N}$ is the ZF equalization matrix (an $N \times N$ diagonal matrix), i.e.,
\begin{equation}\label{EQ_EQ666}
{{{\mathbf{G}}_{\mathrm{ZF}}} = \left[ {\begin{array}{*{20}{c}}
{\frac{1}{{{\widehat{H}_{\mathrm{CE}}}\left( 1 \right)}}}&{}&{}&{}\\
{}&{\frac{1}{{{\widehat{H}_{\mathrm{CE}}}\left( 2 \right)}}}&{}&{}\\
{}&{}& \ddots &{}\\
{}&{}&{}&{\frac{1}{{{\widehat{H}_{\mathrm{CE}}}\left( N \right)}}}
\end{array}} \right]},
\end{equation}
 with ${\widehat{H}_{\mathrm{CE}}(m)}, m = 1,2,\cdots,N$, being the $m$th entry of $\mathbf{\widehat{H}}_{\mathrm{CE}}$. In order to facilitate the data detection in time domain, the time domain output of the ZF equalization, denoted as ${\mathbf{\widehat{s}}_{\mathrm{ZF}}}$, is obtained according to

\begin{equation}\label{EQ_EQ777}
{\mathbf{\widehat{s}}_{\mathrm{ZF}}} = {\mathbf{F}^H_N} {\mathbf{\widehat{S}}_{\mathrm{ZF}}}.
\end{equation}
%

\subsubsection{Refining Data Detection}
After the ZF equalization, the SD-Net is utilized to alleviate the nonlinear distortion by solving a nonlinear problem. The main steps are explained as follows.

\textbf{Constructing SD-Net:} Similar to constructing CE-Net, the neurons and layers of SD-Net are first set by referencing the methods in \cite{Ref2424}. Then, we gradually increase the neurons and layers until the BER can not be reduced significantly to make a trade off between structure complexity and system's BER performance.
The SD-Net, with architecture illustrated in Fig. \ref{fig222}, consists of an input layer, three hidden layers, and an output layer. The network architecture of SD-Net is summarized in Table \ref{table III}, and the detailed descriptions are given as follows.

\begin{table}[t]
\renewcommand{\arraystretch}{1}
\caption{Architecture of SD-Net}
\label{table III} \centering
\setlength{\tabcolsep}{1.4mm}{
\begin{tabular}{c|c|c|c|c|c}
\Xhline{1.2pt}
Layer & Input   & Hidden 1 & Hidden 2 & Hidden 3 & Output \\
\Xhline{1.2pt}
Number of neurons   & ${2N}$  & ${2N}$  & ${12N}$  & ${6N}$ & ${2N}$    \\
\hline
Batch normalization &   $\surd$   &  $\times$    & $\times$    & $\times$    & $\times$ \\

\hline
Activation function & None     & ReLU    & ReLU    & ReLU    & Linear \\

\hline

\end{tabular}}
\end{table}

\begin{itemize}
  \item
  Unlike the CE-Net, the SD-Net needs more neurons for hidden layers to enhance the data detection (including the channel equalization) due to the information discarding in DDST scheme. Based on extensive experiments of network training and validating, the neurons of input layer, output layer, and the first hidden layer are all set as ${2N}$, while ${12N}$ and ${6N}$ neurons are employed for the second hidden layer and the third hidden layer, respectively.
\end{itemize}

\begin{itemize}
  \item
  Similar to the CE-Net, BN is adopted for the input layer of SD-Net, whose inputs are normalized as zero mean and unit variance, to accelerate convergence and prevent overfitting \cite{a9}.
\end{itemize}

\begin{itemize}
  \item
 For the output layer, the linear activation function is adopted, while for the three hidden layers, the activation function ReLUs are adopted.
\end{itemize}


\textbf{Data detection:} Based on the complex-valued ${\mathbf{\widehat{s}}_{\mathrm{ZF}}}$ given in (\ref{EQ_EQ777}), the real-valued form ${\mathbf{\widetilde{s}}_{\mathrm{ZF}}} \in\mathbb{R}^{2N\times1}$ is given by
 \begin{equation}\label{EQ_Szf122}
 {{{\bf{\widetilde{s}}}_{\mathrm{ZF}}} = {{\left[ {{\mathop{\rm Re}\nolimits} \left( {{\bf{\widehat{s}}}_{\mathrm{ZF}}^T} \right),{\mathop{\rm Im}\nolimits} \left( {{\bf{\widehat{s}}}_{\mathrm{ZF}}^T} \right)} \right]}^T}}.
 \end{equation}
The SD-Net is developed to refine the data detection performance by solving a nonlinear problem. By denoting the output of the SD-Net as ${\mathbf{\widetilde{s}}_{\mathrm{SD}}}$, we have
\begin{equation}\label{EQ_EQ999}
{\mathbf{\widetilde{s}}_{\mathrm{SD}}} = {f^{\left( {\cal I} \right)}}\left( {\mathbf{W}_{\mathrm{SD}}^{\left( {\cal I} \right)}{f^{\left( {{\cal I} - 1} \right)}}\left( { \cdots {f^{\left( 2 \right)}}\left( {\mathbf{W}_{\mathrm{SD}}^{\left( 2 \right)}{\mathbf{\widetilde{s}}_{\mathrm{ZF}}} + \mathbf{b}_{\mathrm{SD}}^{\left( 2 \right)}} \right)} \right) + {\mathbf{b}_{\mathrm{SD}}^{\left( {\cal I} \right)}}} \right),
\end{equation}
where $\mathbf{W}_{\mathrm{SD}}^{\left( i  \right)}$ and $\mathbf{b}_{\mathrm{SD}}^{\left( i  \right)}$ are the weight matrix and bias vector of the $i$th layer $\forall i  = 2, \cdots ,{\cal I}$, respectively. Then, the real-valued ${\mathbf{\widetilde{s}}_{\mathrm{SD}}}$ is reshaped to the complex-valued ${\mathbf{\widehat{s}}}$. Compared with the data detection in DDST with MMSE equalization \cite{Ref_4}, the refined ${\mathbf{\widehat{s}}}$ can improve system's BER performance, and this will be verified in Section IV.


%
%
%
%
%
%
%
%
 \textbf{Model training:} The training details of the SD-Net, which include the training data collection, loss function description and network parameters initialization, are explained as follows.
\begin{itemize}
  \item \textit{Data collection:} Similar to the data collection of CE-Net, the basic data generation method is adopted in SD-Net to acquire the training set $\{ {{\mathbf{\widetilde{s}}}_{\mathrm{ZF}}}, {\mathbf{\widetilde{s}}} \}$. For training SD-Net, the trained network parameters of CE-Net, i.e., $\left\{ {{{\mathbf{W}}_{\mathrm{CE}}^{(\ell)}}; {{\mathbf{b}}_{\mathrm{CE}}^{(\ell)}}} \right\}$, $\ell = 2,3,4$, are fixed, and we employ the CE-Net to generate the refined CSI ${\mathbf{\widetilde{H}}}_{\mathrm{CE}}$ (or ${\mathbf{\widehat{H}}}_{\mathrm{CE}}$). According to (\ref{EQ_EQ3333})--(\ref{EQ_EQ777}), the detection symbol ${{\mathbf{\widehat{s}}_{\mathrm{ZF}}}}$ is obtained. Then, we map the complex-valued ${{\mathbf{\widehat{s}}_{\mathrm{ZF}}}}$ and modulated symbol ${ \mathbf{\widetilde{s}} }$ to real-valued formations, and thus form the real-valued ${\{{\mathbf{\widetilde{s}}_{\mathrm{ZF}}}\}}$ and ${\{ \mathbf{\widetilde{s}}\} }$, respectively. Similar to the generation of SD-Net's training set, an SD-Net's validation set is also generated.

\end{itemize}

%
%

\begin{itemize}
\item \textit{Loss function:} The loss function for SD-Net is given by
 \begin{equation}\label{EQ20}
{{{L}_{\mathrm{SD - Net}}} = \frac{1}{{{T_2}}}\left\| {\mathbf{\widetilde{s}} - \mathbf{\widetilde{s}}_{\mathrm{SD}}} \right\|_2^2 + {\alpha _{\mathrm{SD}}}\sum\limits_{i = 2}^5 {{{\left\| {\mathbf{W}_{\mathrm{SD}}^{(i)}} \right\|}_2^2}} },
\end{equation}
where ${T_2}$ is the total number of samples in training set of SD-Net training, ${\alpha _{\mathrm{SD}}}$ denotes the regularization coefficient of the SD-Net. Similar to the optimization of ${\alpha _{\mathrm{CE}}}$ (see (\ref{EQ18}) for details), the regularization coefficient ${\alpha _{\mathrm{SD}}}$ needs to be optimized to improve SD-Net's generalization performance, which is also discussed in Section IV-{B}.
\end{itemize}

\begin{itemize}
  \item \textit{Weight and bias initialization:} Similar to CE-Net, the elements of weight and bias are initialized as i.i.d Glorot uniform distribution for SD-Net. For expression convenience, we use $\left\{ {{{\mathbf{W}}_{\mathrm{SD}}^{(i)}}; {{\mathbf{b}}_{\mathrm{SD}}^{(i)}}} \right\}$, $i = 2,3,4,5$, to represent the set of training parameters.
\end{itemize}

\subsection{Online Running}
With the trained network parameters of CE-Net and SD-Net (by off-line training), the online running procedure is presented in Table \ref{table_V}. Explanations of online running are given as follows.

\begin{table}[t]
\renewcommand\arraystretch{1.2}
\caption{Online Running Procedure}
\label{table_V}\centering
\begin{tabular}{l}
\hline
\hline
\kern -2pt \textbf{Input:} The received signal ${\mathbf{y}}$ and the training sequence ${\mathbf{c}}$
\\
\hline

\kern 4pt 1): According to (\ref{EQ13_add})--(\ref{EQ15_add333}), the LS estimation is performed to obtain ${{\mathbf{\widehat{H}}_{\mathrm{LS}}}}$.\\

\kern 4pt 2): Use the CE-Net to acquire the refined channel estimation ${{\mathbf{\widehat{H}}_{\mathrm{CE}}}}$.\\

\kern 4pt 3): According to (\ref{EQ_EQ3333})--(\ref{EQ_EQ777}), the ZF equalization is implemented to obtain ${\mathbf{\widehat{s}}_{\mathrm{ZF}}}$.\\

\kern 4pt 4): Use the SD-Net to refine the data detection and obtain ${\mathbf{\widehat{s}}}$.\\

\hline

\kern -2pt \textbf{Output:} The detected signal ${\mathbf{\widehat{s}}}$.\\

 \hline
 \hline
\end{tabular}
\end{table}

\subsubsection{Input Requirement}
\
\newline
\indent
For online running, the received signal ${\mathbf{y}}$ (given in (\ref{EQ1})) and the known training sequence $\mathbf{c}$ are needed. To facilitate the signal processing in the frequency domain, ${\mathbf{y}}$ and $\mathbf{c}$ are transformed to the frequency domain according to (\ref{EQ13_add}) to form ${\mathbf{Y}}$ and $\mathbf{C}$, respectively.

\subsubsection{Channel Estimation}
\
\newline
\indent
\textit{LS estimation:} By using ${\mathbf{Y}}$ and ${\mathbf{C}}$ (given in (\ref{EQ13_add})), the LS estimation, in step 1) of Table \ref{table_V}, is performed according to (\ref{EQ13_add})--(\ref{EQ15_add333}). Then, the channel estimation ${{\mathbf{\widehat{H}}_{\mathrm{LS}}}}$ is obtained, and thus forms the network input of CE-Net (i.e., ${{\mathbf{\widetilde{H}}_{\mathrm{LS}}}}$) according to (\ref{EQ16_add111}).

\textit{Channel estimation using CE-Net:} With network input ${{\mathbf{\widetilde{H}}_{\mathrm{LS}}}}$, the CE-Net refines the channel estimation, and thus acquires the refined CSI ${{\mathbf{\widetilde{H}}_{\mathrm{CE}}}}$ with real-valued form (given in (\ref{EQ16_add})). Then, the complex-valued form of refined CSI, i.e., ${{\mathbf{\widehat{H}}_{\mathrm{CE}}}}$, is obtained according to

\begin{equation}\label{CeNet_add}
{\left\{ {\begin{array}{*{20}{l}}
{  {\mathop{\rm \mathrm{Re}}\nolimits}  \left( {{{\mathbf{\widehat{H}}}_{\mathrm{CE}}}} \right) = {{\mathbf{\widetilde{H}}_{\mathrm{LS}}}} (1:N) }\\
{  {\mathop{\rm \mathrm{Im}}\nolimits}  \left( {{{\mathbf{\widehat{H}}}_{\mathrm{CE}}}} \right) = {{\mathbf{\widetilde{H}}_{\mathrm{LS}}}} (N+1:2N) }
\end{array}} \right.}.
\end{equation}
That is, the real part and imaginary part of ${{\mathbf{\widehat{H}}_{\mathrm{CE}}}}$ are formed by extracting the first $N$ entries and the last $N$ entries of ${{\mathbf{\widetilde{H}}_{\mathrm{CE}}}}$, respectively. This procedure of channel estimation by using CE-Net is listed as step 2) of Table \ref{table_V}.

\subsubsection{Data Detection}
\
\newline
\indent
\textit{ZF equalization:} With the estimated ${{\mathbf{\widehat{H}}_{\mathrm{CE}}}}$, the ZF equalization in the frequency domain is performed according to (\ref{EQ_EQ3333})--(\ref{EQ_EQ555}), and thus obtains ${\mathbf{\widehat{S}}_{\mathrm{ZF}}}$. Then, ${\mathbf{\widehat{S}}_{\mathrm{ZF}}}$ is transformed to time-domain to form ${\mathbf{\widehat{s}}_{\mathrm{ZF}}}$ according to (\ref{EQ_EQ777}). This ZF equalization procedure is given in step 3) of Table \ref{table_V}. By utilizing (\ref{EQ_Szf122}), the real-valued ${\mathbf{\widetilde{s}}_{\mathrm{ZF}}}$ is formed based on the complex-valued ${\mathbf{\widehat{s}}_{\mathrm{ZF}}}$ to facilitate the real-valued requirement of SD-Net.

\textit{Data detection using SD-Net:} With the network input ${{\mathbf{\widetilde{s}}_{\mathrm{ZF}}}}$, the SD-Net refines the data detection. Then, the SD-Net outputs the detected symbol ${{\mathbf{\widetilde{s}}_{\mathrm{SD}}}}$ (given in (\ref{EQ_EQ999})). By extracting the first $N$ entries and the last $N$ entries of ${{\mathbf{\widetilde{s}}_{\mathrm{SD}}}}$, the real part and imaginary part of $\mathbf{\widehat{s}}$ are expressed as

\begin{equation}\label{EqNet_add}
{\left\{ {\begin{array}{*{20}{l}}
{  {\mathop{\rm \mathrm{Re}}\nolimits}  \left( {{{\mathbf{\widehat{s}}}}} \right) = {{\mathbf{\widetilde{s}}_{\mathrm{SD}}}} (1:N) }\\
{  {\mathop{\rm \mathrm{Im}}\nolimits}  \left( {{{\mathbf{\widehat{s}}}}} \right) = {{\mathbf{\widetilde{s}}_{\mathrm{SD}}}} (N+1:2N) }
\end{array}} \right.}.
\end{equation}
This data detection by using SD-Net is presented in step 4) of Table \ref{table_V}.

From step 1) to 4) in Table \ref{table_V}, the refined channel estimation and data detection can be achieved from the proposed CE-Net and SD-Net with the received signal ${\mathbf{y}}$ and the known training sequence ${\mathbf{c}}$. Compared with the conventional DDST with MMSE channel estimation and MMSE equalization in \cite{Ref_4}, the proposed scheme can achieve a lower BER at the cost of off-line training time, which is the usual feature of DL \cite{IEERef111}. It is worth noting that, the performance of the proposed scheme is improved without any second-order statistic information of wireless channel $\mathbf{h}$ and noise $\mathbf{v}$.

\section{Simulation Results}

In this section, we present the numerical results for the proposed receiver scheme. The basic parameters and definitions involved in simulations are first given in Section IV-{A}. Then, in Section IV-{B}, the model training is analyzed by discussing the regularization coefficient and the training signal-to-noise ratio (SNR). The effectiveness of  the proposed receiver scheme is verified in Section IV-{C}. At last, the analysis of parameters robustness are given in Section IV-{D}.

\subsection{Parameter Setting}
In the experiments, the following basic parameters are applied unless otherwise specified. From \cite{Ref_4}, $N=240$, $P=12$, and $t=0$ are considered. The channel $\mathbf{h}$ defined in (\ref{EQ1}) is generated by channel model COST2100 \cite{a7_add} with outdoor semi-urban scenario at the 300MHz band, and the number of multi-path is set as $L=12$. The transmitted data symbol ${\mathbf{{s}}}$ given in (\ref{EQ4}) is modulated by QPSK modulation. The training and validation sets of both CE-Net and SD-Net have 60,000 and 20,000 samples, respectively. Their batch sizes are all set as 80 samples. We use Adam optimizer as the training optimization algorithm \cite{IV-1} with parameters ${\beta _1} = 0.99$ and ${\beta _2} = 0.999$ \cite{IV-2}. The learning rate is set to 0.0001, and the $L_2$ regularization \cite{IV-3} is adopted.
The SNR in decibel (dB) is defined as
\begin{equation}\label{SNR35_add}
{\mathrm{SNR} = 10{\log _{10}}\left( {\frac{{{E_{\mathbf{x}_{\mathrm{dis}}}}}}{{\sigma _v^2}}} \right)},
\end{equation}
where ${E_{\mathbf{x}_{\mathrm{dis}}}}$ is the transmitted power of ${\mathbf{x}_{\mathrm{dis}}}$, which is equal to the summation of data-symbol power ${E_s}$ and training-sequence power ${E_c}$. In these simulations, ${E_s} = 0.9{\mathbf{x}_{\mathrm{dis}}}$ and ${E_c} = 0.1{\mathbf{x}_{\mathrm{dis}}}$.
For network training, the mixed SNR is adopted, i.e., each training sample is generated under a random SNR from $\mathrm{SNR}=0$dB to $\mathrm{SNR}=45$dB with the interval of 5dB.

For hardware imperfection, the effects of nonlinear distortion caused by HPA are considered in this paper. From \cite{a14}, the nonlinear amplitude and phase are given by
\begin{equation}\label{EQ_nonlinear}
A\left( x \right) = \frac{{{\alpha _a}x}}{{1 + {\beta _a}{x^2}}},~{\rm{ }}\Phi \left( x \right) = \frac{{{\alpha _\phi }{x^2}}}{{1 + {\beta _\phi }{x^2}}},
\end{equation}
where ${\alpha _a} = 1.96$, ${\beta _a} = 0.99$, ${\alpha _\phi } = 2.53$, and ${\beta _\phi } = 2.82$ are considered in the simulations.

Furthermore, we use error  vector  magnitude (EVM) to measure distortion intensity, which is defined as \cite{IV-5}
\begin{equation}\label{SNR36_add}
{{\mathrm{EVM}} = \sqrt {\frac{{\sum\limits_{n \in N} {{{\left| {{{\widetilde x}_n} - {R_n}} \right|}^2}} }}{{\sum\limits_{n \in N} {{{\left| {{R_n}} \right|}^2}} }}}  \times 100\% },
\end{equation}
where ${\widetilde{x}_n}$ and ${R_n}$ are the distorted and ideal undistorted outputs of HPA, respectively. That is, for the same input, ${\widetilde{x}_n}$ and ${R_n}$ are the outputs when HPA works in saturated region and ideal linear region, respectively. Except for the robustness analysis against EVMs, we set the EVM as ${55\%}$.

The training and testing of proposed method are carried out on a server with NVIDIA TITAN RTX GPU and Intel Xeon(R) E5-2620 CPU 2.1GHz$\times$16, and the results of DDST scheme are obtained by using MATLAB simulation on the server CPU due to the lack of a GPU solution.

For the convenience of expression, the simplified expressions in the simulations are given as follows.
\begin{itemize}
  \item ``Proposed'' is used to denote the proposed receiver scheme, i.e., the CE-Net and SD-Net are jointly applied.
  \item ``LS\_CE + ZF\_SD'', ``LS\_CE + MMSE\_SD'', ``MMSE\_CE + ZF\_SD'', and ``MMSE\_CE + MMSE\_SD'' represent ``LS channel estimation followed by ZF equalization'', ``LS channel estimation followed by MMSE equalization'', ``MMSE channel estimation followed by ZF equalization'', and ``MMSE channel estimation followed by MMSE equalization'' in the conventional DDST scheme, respectively. We utilize ``LS\_CE + ZF\_SD'' and ``MMSE\_CE + MMSE\_SD'' as the baselines of the low computational complexity and high detection performance of DDST, respectively.
  \item ``CE-Net + ZF\_SD'', ``CE-Net + MMSE\_SD'', ``LS\_CE + SD-Net'', and ``MMSE\_CE + SD-Net'' stand for the ``proposed CE-Net followed by ZF equalization'', ``proposed CE-Net followed by MMSE equalization'', ``LS channel estimation followed by the proposed SD-Net'', and ``MMSE channel estimation followed by the proposed SD-Net'', respectively.
\end{itemize}


\subsection{Regularization Coefficient and Training SNR Optimization}
Usually, the model training is influenced by regularization coefficient and training SNR \cite{Ref803_1, a5}. To obtain the optimized learning parameters, the impacts of regularization coefficient and training SNR are analysed by using the following simulation experiments.
\subsubsection{Regularization Coefficient}

\begin{figure*}[t]
\centering
\includegraphics[scale=0.65]{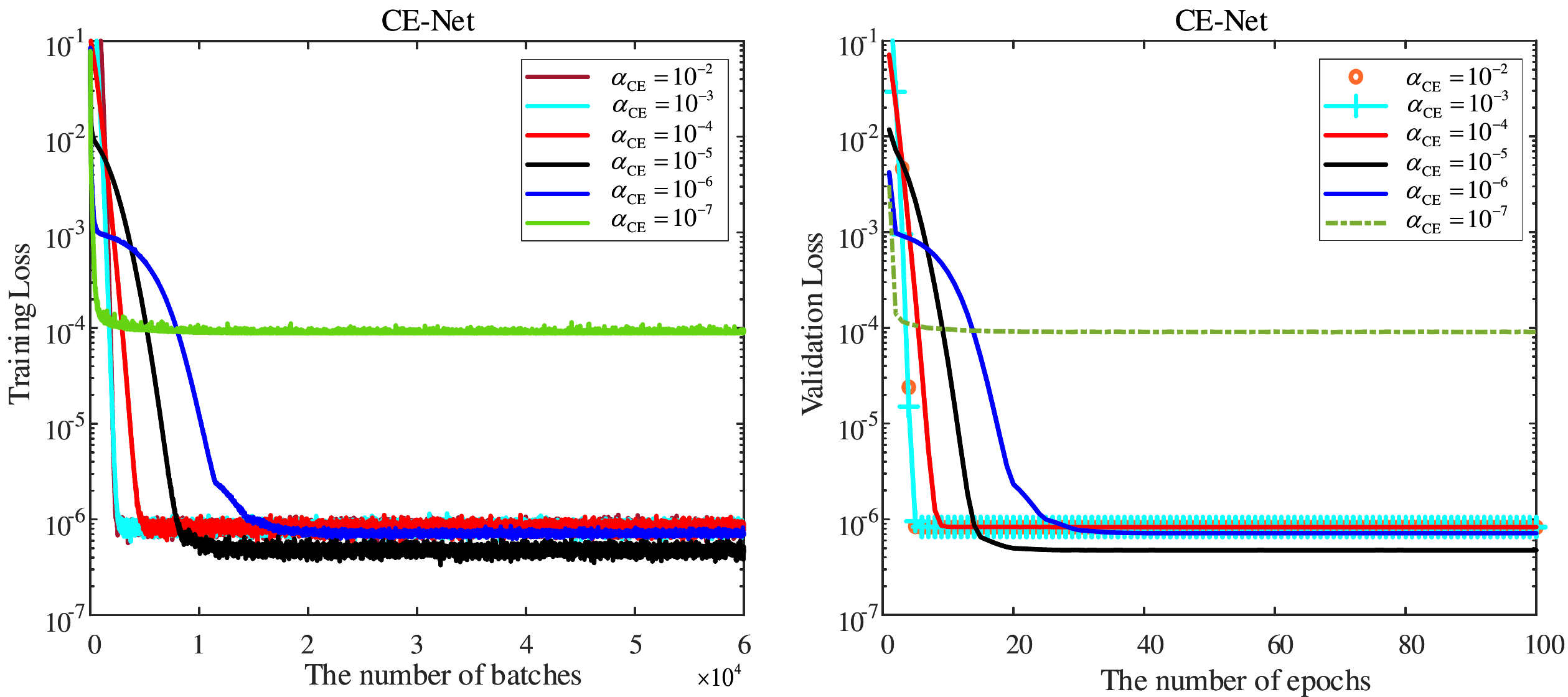}
\caption{ Training and validation losses against the impact of regularization coefficient of CE-Net.}
\label{figCELoss}
\end{figure*}

\begin{figure*}[t]
\centering
\includegraphics[scale=0.65]{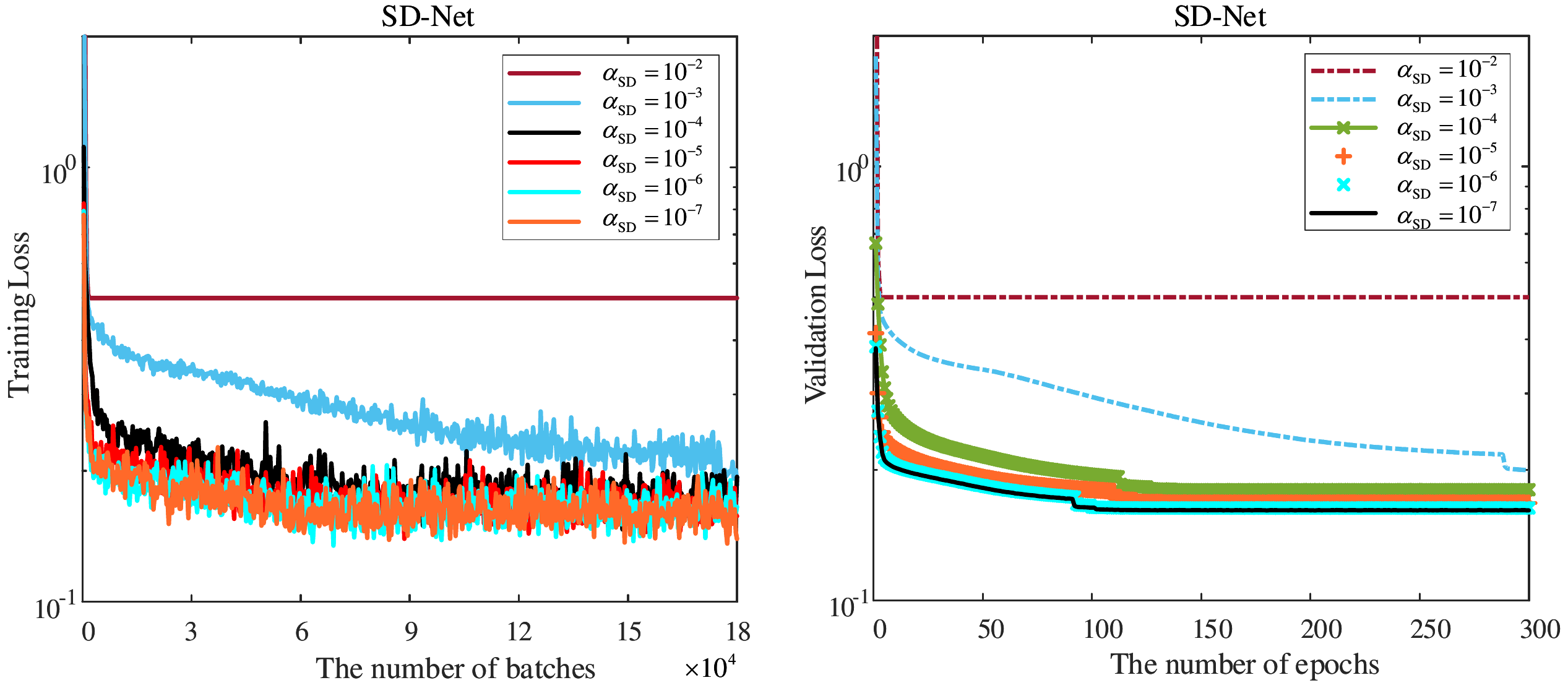}
\caption{ Training and validation losses against the impact of regularization coefficient of SD-Net.}
\label{figSDLoss}
\end{figure*}

From (\ref{EQ18}) and (\ref{EQ20}), the $L_2$ regularization is employed to CE-Net and SD-Net. To capture an optimized regularization coefficient, the training and validation losses are given in the Fig. \ref{figCELoss} and Fig. \ref{figSDLoss} for CE-Net and SD-Net, respectively.

In Fig. \ref{figCELoss}, different values of ${\alpha _{\mathrm{CE}}}$ (i.e., ${\alpha _{\mathrm{CE}}} = {10^{ - 2}},{10^{ - 3}}, \ldots ,{10^{ - 7}}$) are employed for CE-Net, where the numbers of the training batches and validation epochs are set as ${\rm{6}} \times {\rm{1}}{{\rm{0}}^{\rm{4}}}$ and 100, respectively. From Fig. \ref{figCELoss}, it could be observed that different regularization coefficients achieve different convergence values for the training or validation loss. Even so, the convergence values of the training and validation losses are almost the same for a given ${\alpha _{\mathrm{CE}}}$, which shows the CE-Net possesses a good generalization performance. There exists a weak correlation between the convergence rate and the convergence value. Among the given values of ${\alpha _{\mathrm{CE}}}$, the fastest convergence rate of training (or validation) loss is observed when ${\alpha _{\mathrm{CE}}} = {10^{ - 7}}$, while reaching the maximum convergence value. On the contrary, the slowest convergence rate of training (or validation) loss is obtained when ${\alpha _{\mathrm{CE}}} = {10^{ - 6}}$, but its convergence value is not the minimum one. From the plotted curves in Fig. \ref{figCELoss}, the smallest convergence value is achieved when ${\alpha _{\mathrm{CE}}} = {10^{ - 5}}$.

In Fig. \ref{figSDLoss}, to harvest an optimized regularization coefficient, different values of ${\alpha _{\mathrm{SD}}}$ (i.e., ${\alpha _{\mathrm{SD}}} = {10^{ - 2}},{10^{ - 3}}, \ldots ,{10^{ - 7}}$) are tested in the SD-Net. The numbers of the training batches and validation epochs are set as ${\rm{1.8}} \times {\rm{1}}{{\rm{0}}^{\rm{5}}}$ and 300, respectively. From Fig. \ref{figSDLoss}, for each given regularization coefficient ${\alpha _{\mathrm{SD}}}$, the convergence values of the training and validation losses are almost the same, which indicates the good generalization performance of the SD-Net. In addition, when $\alpha_{\mathrm{SD}} $ is set as $10^{-7}$, the convergence value of training (or validation) loss is the minimum one while its convergence rate is neither the fastest nor the slowest (the fastest and slowest convergence rates are obtained by  $\alpha_{\mathrm{SD}} = 10^{-2}$ and $\alpha_{\mathrm{SD}} = 10^{-3}$, respectively). This phenomenon shows a weak correlation between the convergence rate and the convergence value. Even so, the minimum convergence value of validation loss is still observed with high probability when $\alpha_{\mathrm{SD}} = 10^{-7}$.

In this paper, to obtain a good performance (e.g., the estimation accuracy, detection rightness, and generalization, etc), we choose the regularization coefficients according to the minimum validation loss with high probability, thereby $\alpha_{\mathrm{CE}} = 10^{-5}$ and $\alpha_{\mathrm{SD}} = 10^{-7}$ are employed for CE-Net and SD-Net, respectively.
\begin{figure}[t]
\centering
\includegraphics[scale=0.705]{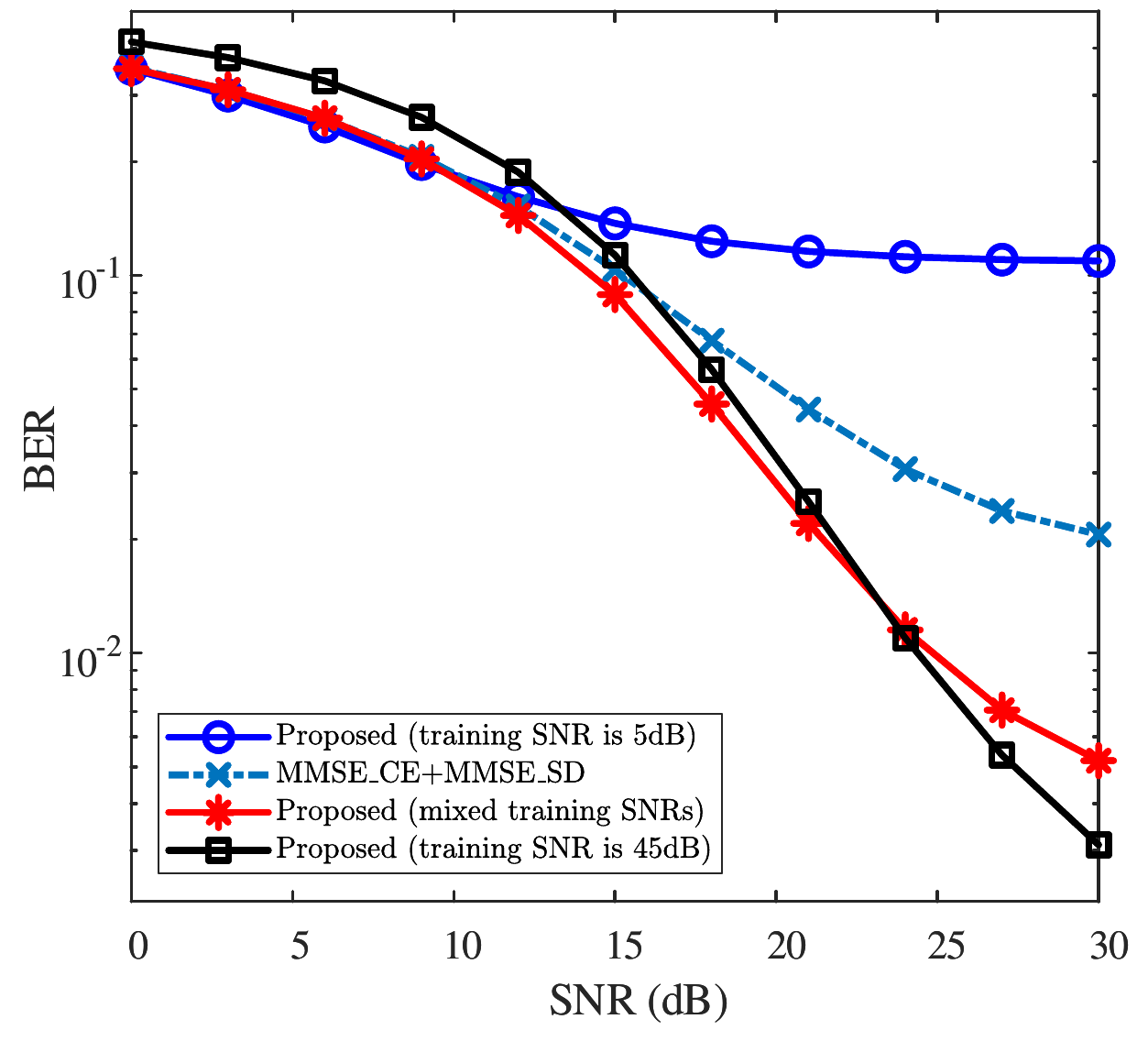}
\caption{ BER performances against the impact of training SNR, where $N = 240$, $P = 12$, $L = 12$, and $\mathrm{EVM} = 55\%$.}
\label{fig5}
\end{figure}
\subsubsection{Training SNR}
\text{\\ \\ \\}
\begin{figure}[t]
\centering
\includegraphics[scale=0.702]{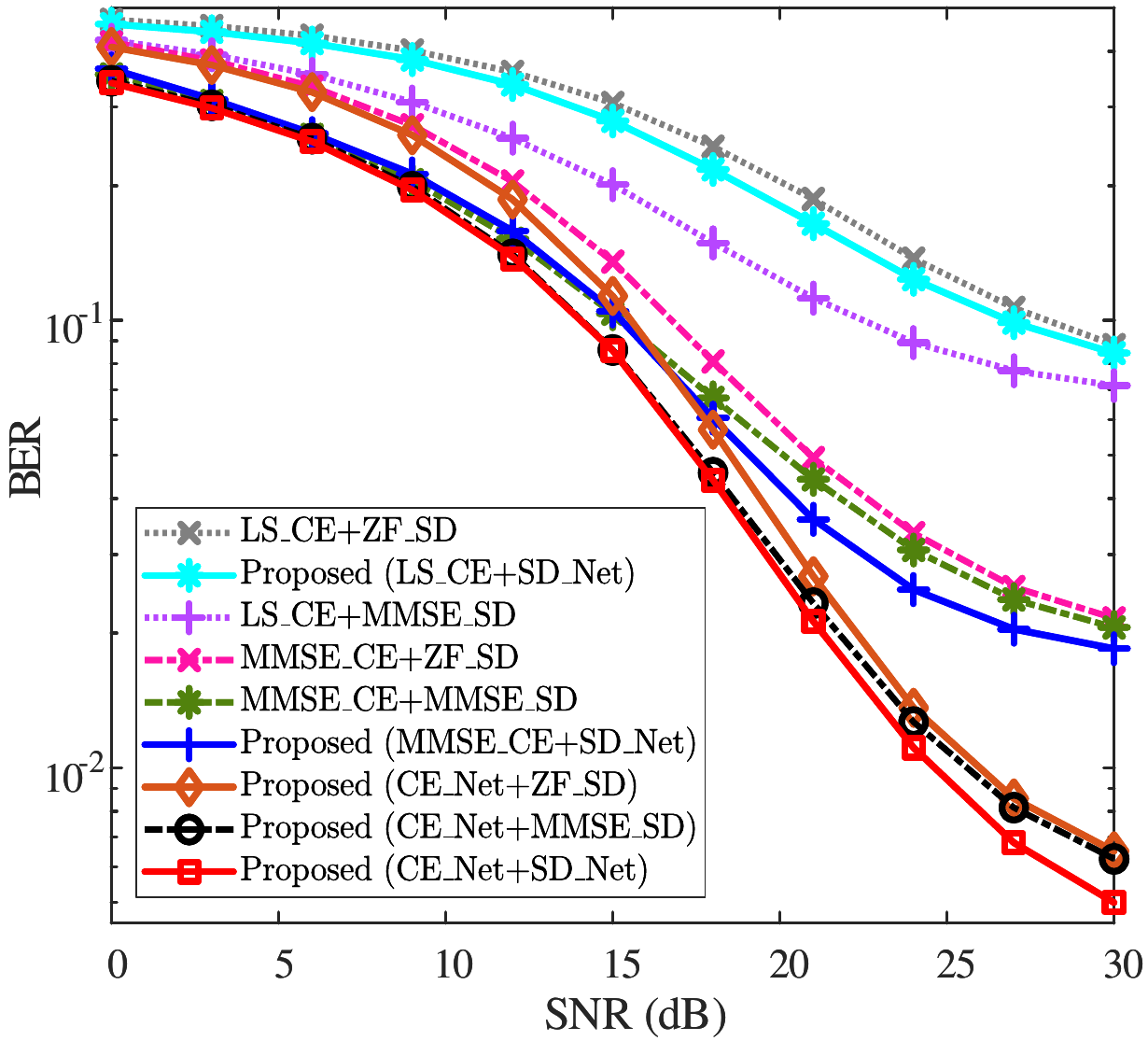}
\caption{BER performance for different methods vs. SNR, where $N = 240$, $P = 12$, $L = 12$, and $\mathrm{EVM} = 55\%$.}
\label{fig4}
\end{figure}
As mentioned in \cite{a5}, the training SNR usually causes DL network's performance fluctuation. To analyze the impact of training SNR on model training, the BER curves with different training SNRs are plotted in Fig. \ref{fig5}.

In Fig. \ref{fig5}, ${{\mathrm{SNR}} = 5}$dB (low training SNR), mixed SNR (each training sample is generated under a random SNR from ${{\mathrm{SNR}} = 0}$dB to ${{\mathrm{SNR}} = 45}$dB with the interval of 5dB), and ${{\mathrm{SNR}} = 45}$dB (high training SNR) are considered as the training SNRs. For training SNR being $5$dB, the ``Proposed'' captures similar BER with ``MMSE\_CE + MMSE\_SD'' in the relatively low testing SNR region (e.g., ${{\mathrm{SNR}} \leq 10}$dB), while appearing much higher BER than that of ``MMSE\_CE + MMSE\_SD'' in the relatively high testing SNR region (e.g., ${{\mathrm{SNR}} \geq 20}$dB). It seems that the low training SNR is unsuitable for training the proposed network worked in the high SNR region. Although the high training SNR can bring a low BER when the testing SNR is larger than 25dB, it cannot reap a satisfied BER in the low SNR region (e.g., the testing SNR is lower than 10dB). Hence, the training $\mathrm{SNR}=45$dB is also unsuitable for the proposed scheme. When mixed SNR is adopted as the training SNR, the BER of ``Proposed'' is similar or lower than that of the ``MMSE\_CE + MMSE\_SD'' for the testing SNRs varying from 0dB to 30dB. From Fig. \ref{fig5}, both the low training SNR and the high training SNR have their drawbacks, especially for the case where the testing SNR is far from the training SNR. Thus, we employ the mixed SNR as the training SNR for a trade-off consideration.

\subsection{Analysis of Simulation Performance}
 To verify the effectiveness of the proposed receiver scheme, the BER performance of the proposed receiver is compared against that of the conventional DDST scheme in \cite{Ref_4}. The performance curves of BER are given in Fig. \ref{fig4}, and partial numerical results are presented in TABLE \ref{table VI} for the convenience of comparison.

\begin{table*}[t]
\renewcommand{\arraystretch}{1}
\caption{The BER of Each Method}
\label{table VI} \centering
\setlength{\tabcolsep}{1.5mm}
{
\begin{tabular}{c|c c c c c c c c c c c}
\hline
\hline
\diagbox[width=13em,dir=SE]{Method}{SNR}  & 0dB & 3dB & 6dB & 9dB & 12dB & 15dB   & 18dB & 21dB & 24dB & 27dB & 30dB \\
\hline
\hline
LS\_CE + ZF\_SD       & 0.470& 0.455 & 0.433& 0.401& 0.359& 0.306& 0.2447 & 0.1869 & 0.1375 & 0.1067 & 0.0881 \\
\hline
LS\_CE + SD\_Net      &0.459&	0.441&	0.417&	0.383& 0.337&	0.279&	0.2177&	0.1647&	0.1237&	0.0989&	0.0846 \\
\hline
LS\_CE + MMSE\_SD     & 0.423 & 0.392& 0.355& 0.307 & 0.255& 0.201 & 0.1491& 0.1122& 0.0893 &0.0772 & 0.0715\\
\hline
MMSE\_CE + ZF\_SD    & 0.419 & 0.381& 0.333  & 0.273& 0.204& 0.135  &0.0810& 0.0493& 0.0335  &0.0254 & 0.0216\\
\hline
MMSE\_CE + MMSE\_SD &0.355 &0.309& 0.259 & 0.206 &0.152 & 0.103 & 0.0672 & 0.0442 &0.0307 &0.0238 & 0.0206\\
\hline
MMSE\_CE + SD\_Net  & 0.365& 0.312& 0.261 &0.212 &0.158 & 0.105  &0.0605  &0.0359  &0.0250 & 0.0204 &  0.0185\\
\hline
CE\_Net + ZF\_SD   & 0.408 & 0.372& 0.324 & 0.259& 0.186& 0.113& 0.0570& 0.0268& 0.0136 & 0.0085 & 0.0065\\
\hline
CE\_Net + MMSE\_SD  & 0.343 &0.302& 0.255 &0.199& 0.140 & 0.085 & 0.0457& 0.0233& 0.0127 & 0.0081 & 0.0063\\
\hline
\textbf{CE\_Net + SD\_Net} & \textbf{0.342} & \textbf{0.301}& \textbf{0.255}&\textbf{ 0.198}& \textbf{0.140}&\textbf{ 0.084 }& \textbf{0.0456}&\textbf{0.0221}& \textbf{0.0115}& \textbf{0.0071}& \textbf{0.0049}\\
\hline
\end{tabular}}
\end{table*}

From Fig. \ref{fig4} and TABLE \ref{table VI}, the ``Proposed'' achieves the minimum BER. That is, the BER performance of ``Proposed'' outperforms those of ``MMSE\_CE + SD\_Net'', ``CE\_Net + ZF\_SD'', ``CE\_Net + MMSE\_SD'', and ``MMSE\_CE + MMSE\_SD'', especially in the relatively high SNR region (e.g., ${{\mathrm{SNR}} \geq 20}$dB). In the relatively low SNR region (e.g., ${{\mathrm{SNR}} \leq 5}$dB), the ``Proposed'' has similar BER performance as those of ``MMSE\_CE + SD\_Net'', ``CE\_Net + ZF\_SD'', ``CE\_Net + MMSE\_SD'', and ``MMSE\_CE + MMSE\_SD''.  The ``LS\_CE + ZF\_SD'' has the worst BER performance due to the lack of channel and noise's second-order statistics. As a whole, even compared with ``MMSE\_CE + MMSE\_SD'', the proposed receiver scheme achieves a lower BER, while avoiding the requirements of channel and noise's second-order statistics.

In addition, the effectiveness of the proposed CE-Net and SD-Net is also verified in Fig. \ref{fig4}. For CE-Net, the BERs of ``CE\_Net + ZF\_SD'' and ``CE\_Net + MMSE\_SD'' are not higher than those of ``MMSE\_CE + ZF\_SD'' and ``MMSE\_CE + MMSE\_SD'' in all given SNR regions. In particular, in the relatively high SNR region, the BER of ``CE\_Net + ZF\_SD'' or ``CE\_Net + MMSE\_SD'' is lower than those of ``MMSE\_CE + ZF\_SD'' and ``MMSE\_CE + MMSE\_SD''. Thus, the proposed CE-Net has its effectiveness in improving the accuracy of channel estimation. For the SD-Net, the BER of the ``LS\_CE + SD\_Net'' is lower than that of ``LS\_CE + ZF\_SD'' in all SNR regions, and the ``MMSE\_CE + SD\_Net'' obtains a lower BER than ``MMSE\_CE + MMSE\_SD'' when ${{\mathrm{SNR}} > 15}$dB while maintaining similar BER for ${{\mathrm{SNR}} \leq 15}$dB. These reflect that the proposed SD-Net improves the BER performance of DDST without the information of noise's second-order statistics.

\begin{figure*}[t]
\centering
\includegraphics[scale=0.46]{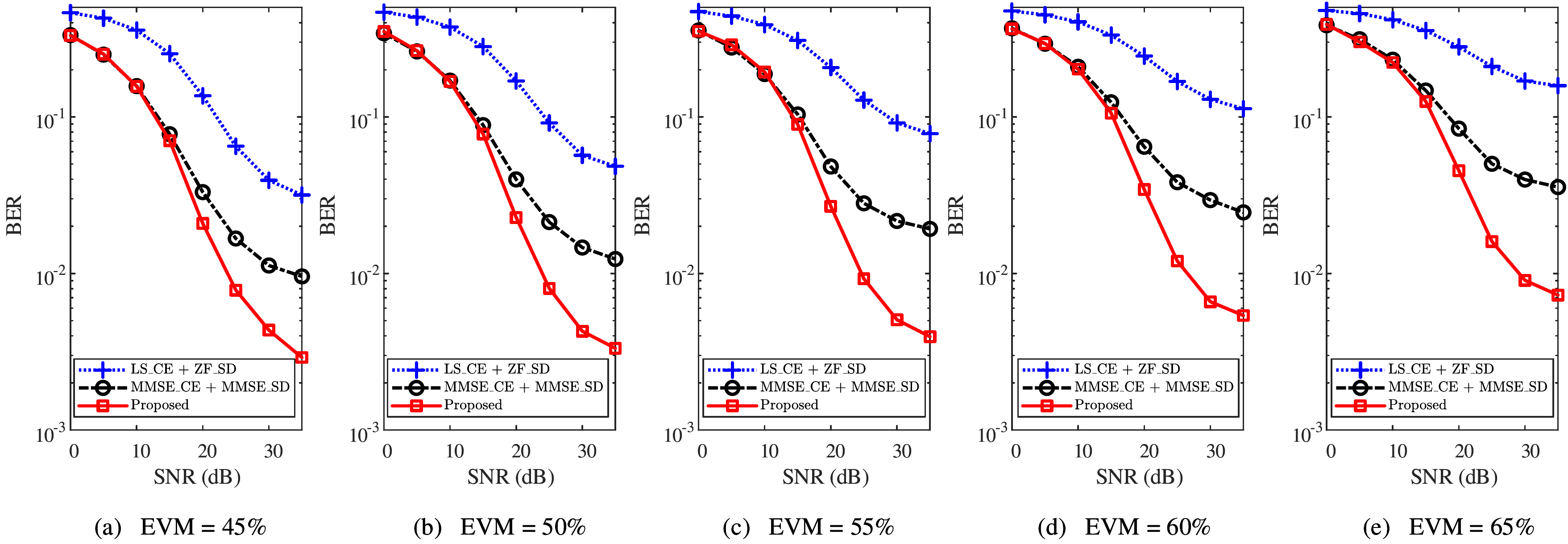}
\caption{BER performance against the impact of EVM, where $\mathrm{EVM} = 45\%$, $\mathrm{EVM} = 50\%$, $\mathrm{EVM} = 55\%$, $\mathrm{EVM} = 60\%$, and $\mathrm{EVM} = 65\%$ are considered, respectively.}
\label{fig6}
\end{figure*}

\subsection{Analysis of Parameter Impact}

In this subsection, the robustness of the proposed receiver scheme against the impacts of EVM and $L$ is analysed. For the ease of analysis, only one impact parameter is changed while other basic parameters are freezed during the simulations.

\subsubsection{Impact of EVM}

Usually, the signals with different amplitudes cause different distortion degrees when they pass through the same HPA working in the saturated region. In order to reflect the robustness of the proposed receiver scheme, we analyze the parameter impact against EVM in Fig. \ref{fig6}.

In Fig. \ref{fig6}, the EVM varies from $45\%$ to $65\%$ with the interval of $5\%$. For each given EVM, the ``Proposed'' achieves the minimum BER compared with ``LS\_CE + ZF\_SD'' and ``MMSE\_CE + MMSE\_SD''. This reflects that the ``Proposed'' can improve the BER performance of DDST with little impact from EVM. In particular, for the relatively high SNR region, e.g., ${{\mathrm{SNR}} \geq 20}$dB, the BER of ``Proposed'' is significantly lower than those of ``LS\_CE + ZF\_SD'' and ``MMSE\_CE + MMSE\_SD''. In addition, the improvement of BER performance can be observed for each given EVM. Thus, the proposed receiver scheme can significantly improve the BER performance of DDST scheme, and this improvement is robust to the EVM variations.

\begin{figure*}[t]
\centering
\includegraphics[scale=1.11]{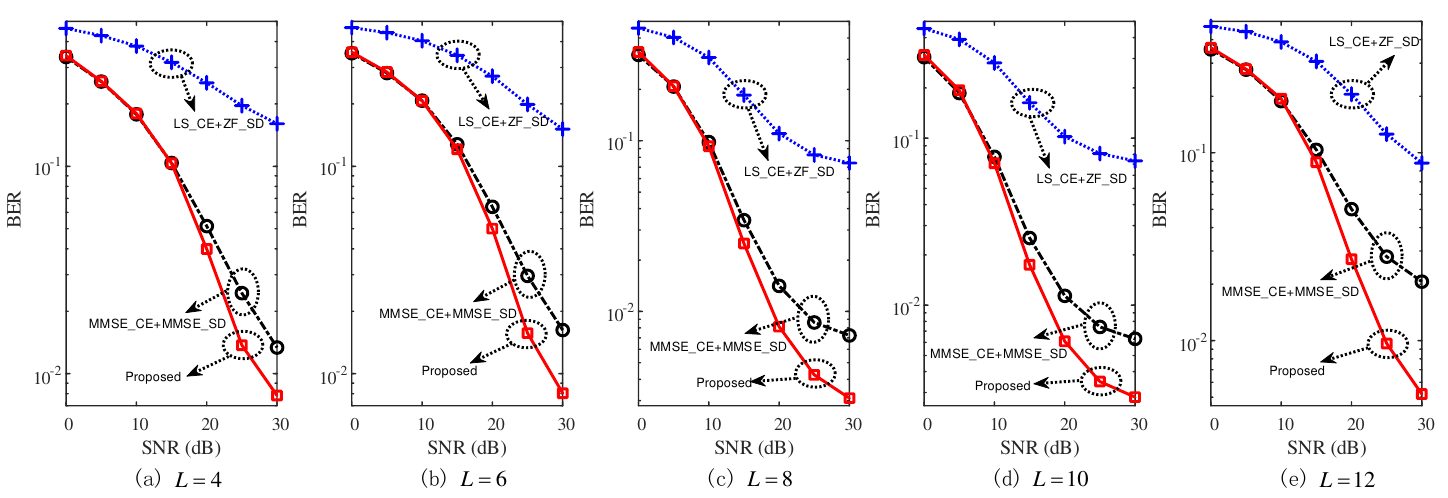}
\caption{BER performance against the impact of $L$, where $L = 4$, $L = 6$, $L = 8$, $L = 10$, and $L = 12$ are considered, respectively.}
\label{fig7}
\end{figure*}

\subsubsection{Impact of $L$}

The BER performance is usually influenced by the number of multi-path, i.e., ${L}$. To illuminate the robustness against multi-path impact, the BER performance is tested with different values of $L$ in Fig. \ref{fig7}, where $L = 4$, $L = 6$, $L = 8$, $L = 10$, and $L = 12$ are considered.

From Fig. \ref{fig7}, the ``Proposed'' achieves the minimum BER for each given ${L}$ compared with ``LS\_CE + ZF\_SD'' and ``MMSE\_CE + MMSE\_SD''. This reflects that the proposed receiver scheme can improve the BER performance of DDST with ``LS\_CE + ZF\_SD'' or ``MMSE\_CE + MMSE\_SD''. Especially in the relatively high SNR region, e.g., ${{\mathrm{SNR}} \geq 20}$dB, the BER of ``Proposed'' is significantly lower than those of ``LS\_CE + ZF\_SD'' and ``MMSE\_CE + MMSE\_SD''. This verifies that the proposed receiver scheme possesses its robustness against the impact of $L$.

\section{Conclusion}
In this paper, a joint model and data driven receiver scheme for DDST is proposed to alleviate the symbol misidentification caused by hardware imperfection and unsatisfactory recovery of received symbols. This scheme exploits the advantages of linear receivers and nonlinear solutions to tackle nonlinearity. Specifically, the proposed receiver scheme first adopts the model driven modes, employing LS estimation and ZF equalization to highlight the initial features of linear receivers. Then, based on the obtained features, two shallow neural networks, CE-Net and SD-Net are constructed with data driven mode to refine the channel estimation and data detection, respectively. Compared with existing DDST schemes with MMSE based channel estimation and equalization, the CE-Net and SD-Net in the proposed receiver achieve similar or better data detection performance without the second-order statistics of the channel and noise. In addition to the effectiveness in alleviating the hardware imperfection and maintaining the BER performance against parameter variations, the proposed receiver scheme avoids the challenge of mobile device modification in DDST systems and promotes the existing researches move towards practical application.

\appendices

\ifCLASSOPTIONcaptionsoff
  \newpage
\fi



%

\nocite{*}
\bibliographystyle{IEEEtran}
\bibliography{refer}

\begin{thebibliography}{10}

\bibitem{Ref_4}
M.~Ghogho, D.~McLernon, E.~Alameda-Hernandez, and A.~Swami, ``Channel
  estimation and symbol detection for block transmission using data-dependent
  superimposed training,'' \emph{IEEE Signal Process. Lett.}, vol.~12, no.~3,
  pp. 226--229, Mar. 2005.

\bibitem{rr8}
S.~He and J.~K. Tugnait, ``Self-interference suppression in doubly-selective
  channel estimation using superimposed training,'' in \emph{Proc. IEEE Int.
  Conf. Commun.}, Glasgow, UK, June 2007, pp. 3028--3033.

\bibitem{rr10}
H.~Zhang and B.~Sheng, ``An enhanced partial-data superimposed training scheme
  for {OFDM} systems,'' \emph{IEEE Commun. Lett.}, vol.~24, no.~8, pp.
  1804--1807, May 2020.

\bibitem{rr9}
C.~He, G.~Huang, J.~Gao, G.~Dou, and W.~Ying, ``Semiblind channel estimation
  and symbol detection for block transmission using superimposed training,'' in
  \emph{Proc. IEEE Int. Conf. Comp. Inf. Technol}, Chengdu, China, Dec. 2012,
  pp. 627--630.

\bibitem{rr6}
P.~Wang, P.~Fan, W.~Yuan, and M.~Darnell, ``Data detection and coding for
  data-dependent superimposed training,'' \emph{IET Signal Process.}, vol.~8,
  no.~2, pp. 38--145, Apr. 2014.

\bibitem{RefZZ_1}
G.~Dou, C.~Li, J.~Gao, and F.~Guo, ``Constellation rotation and symbol
  detection for data-dependent superimposed training,'' \emph{Electron. Lett.},
  vol.~50, no.~25, pp. 1939--1940, Dec. 2014.

\bibitem{RefZZ_2}
C.~Chan, W.~Huang, C.~Li, and H.~Li, ``Elimination of data identification
  problem for data-dependent superimposed training,'' \emph{IEEE Trans. Signal
  Process.}, vol.~63, no.~6, pp. 1595--1604, Mar. 2015.

\bibitem{rr5}
C.~Kuei, H.~Wei, L.~Chih, and L.~Hsueh, ``Investigation on data identification
  problem for data-dependent superimposed training,'' in \emph{Proc. IEEE Veh.
  Technol. Conf.}, Yokohama, Japan, Jul. 2012, pp. 1--5.

\bibitem{Ref_1}
C.~Kuei, C.~Li, C.~Hung, and W.~Huang, ``A precoding scheme for eliminating
  data identification problem in single carrier system using data-dependent
  superimposed training,'' \emph{IEEE Access}, vol.~7, pp. 45\,930--45\,939,
  Apr. 2019.

\bibitem{I-a3}
X.~Dai, H.~Zhang, and D.~Li, ``Linearly time-varying channel estimation for
  {MIMO/OFDM} systems using superimposed training,'' \emph{IEEE Trans.
  Commun.}, vol.~58, no.~2, pp. 681--693, Feb. 2010.

\bibitem{r4}
T.~Whitworth, M.~Ghogho, and D.~C. McLernon, ``Data identifiability for
  data-dependent superimposed training,'' in \emph{Proc. IEEE Int. Conf.
  Commun.}, Glasgow, UK, June 2007, pp. 2545--2550.

\bibitem{Ref_3}
C.~Qing, W.~Yu, B.~Cai, J.~Wang, and C.~Huang, ``{ELM}-based frame
  synchronization in burst-mode communication systems with nonlinear
  distortion,'' \emph{IEEE Wireless Commun. Lett.}, vol.~9, no.~6, pp.
  915--919, June 2020.

\bibitem{Ref722_1}
C.~Fager, K.~Hausmair, K.~Buisman, K.~Andersson, E.~Sienkiewicz, and
  D.~Gustafsson, ``Analysis of nonlinear distortion in phased array
  transmitters,'' in \emph{Proc. INMMiC}, Graz, Austria, Apr. 2017, pp. 1--4.

\bibitem{Ref722_3}
D.~Korpi, Y.~Choi, T.~Huusari, L.~Anttila, S.~Talwar, and M.~Valkama,
  ``Adaptive nonlinear digital self-interference cancellation for mobile inband
  full-duplex radio: Algorithms and {RF} measurements,'' in \emph{Proc. IEEE
  Global Commun. Conf.}, San Diego, CA, USA, Dec. 2015, pp. 1--7.

\bibitem{Ref722_4}
R.~Dinis, P.~Silva, and T.~Araujo, ``Turbo equalization with cancelation of
  nonlinear distortion for {CP}-assisted and zero-padded {MC-CDM} schemes,''
  \emph{IEEE Trans. Commun.}, vol.~57, no.~8, pp. 2185--2189, Aug. 2009.

\bibitem{Conf_Zjiao}
D.~W. Chi, M.~Al, and P.~Das, ``Effects of channel estimation error and
  nonlinear {HPA} on the performance of {OFDM} in rayleigh channels with
  application to 802.11n {WLAN},'' in \emph{Proc. IEEE Wireless Commun.
  Networking Conf.}, Las Vegas, NV, USA, Apr. 2008, pp. 852--857.

\bibitem{Ref2_Conf25}
G.~Lu, T.~Sakamoto, A.~Chiba, and T.~Kawanishi, ``Experiment investigation of
  nonlinear distortion in {QAM} constellation due to imbalance in intradyne
  coherent receiver,'' in \emph{Proc. Opto-Electron. Commun. Conf.}, Busan
  Korea, Jul. 2012, pp. 339--340.

\bibitem{c9}
H.~Ye, G.~Y. Li, and B.~Juang, ``Power of deep learning for channel estimation
  and signal detection in {OFDM} systems,'' \emph{IEEE Wireless Commun. Lett.},
  vol.~7, no.~1, pp. 114--117, Feb. 2018.

\bibitem{TWC_1}
C.~Fan, X.~Yuan, and Y.~Zhang, ``{CNN}-based signal detection for banded linear
  systems,'' \emph{IEEE Trans. Wireless Commun.}, vol.~18, no.~9, pp.
  4394--4407, June 2019.

\bibitem{c7}
H.~Huang, Y.~Song, J.~Yang, G.~Gui, and F.~Adachi, ``Deep-learning-based
  millimeter-wave massive {MIMO} for hybrid precoding,'' \emph{IEEE Trans. Veh.
  Technol.}, vol.~68, no.~3, pp. 3027--3032, Mar. 2019.

\bibitem{c10}
C.~Qing, B.~Cai, Q.~Yang, J.~Wang, and C.~Huang, ``Deep learning for {CSI}
  feedback based on superimposed coding,'' \emph{IEEE Access}, vol.~7, pp.
  93\,723--93\,733, Jul. 2019.

\bibitem{TWC_2}
J.~Guo, C.~Wen, S.~Jin, and G.~Y. Li, ``Convolutional neural network-based
  multiple-rate compressive sensing for massive {MIMO CSI} feedback: design,
  simulation, and analysis,'' \emph{IEEE Trans. Wireless Commun.}, vol.~19,
  no.~4, pp. 2827--2840, Jan. 2020.

\bibitem{c11}
J.~Yuan, H.~Q. Ngo, and M.~Matthaiou, ``Machine learning-based channel
  prediction in massive {MIMO} with channel aging,'' \emph{IEEE Trans. Wireless
  Commun.}, vol.~19, no.~5, pp. 2960--2973, Feb. 2020.

\bibitem{c4}
H.~Huang, J.~Yang, H.~Huang, Y.~Song, and G.~Gui, ``Deep learning for
  super-resolution channel estimation and {DOA} estimation based massive {MIMO}
  system,'' \emph{IEEE Trans. Veh. Technol.}, vol.~67, no.~9, pp. 8549--8560,
  Sep. 2018.

\bibitem{a8}
J.~Guerreiro, R.~Dinis, and P.~Montezuma, ``Analytical performance evaluation
  of precoding techniques for nonlinear massive {MIMO} systems with channel
  estimation errors,'' \emph{IEEE Trans. Commun.}, vol.~66, no.~4, pp.
  1440--1451, Apr. 2018.

\bibitem{Ref4_24}
C.~Qing, W.~Yu, S.~Tang, C.~Rao, and J.~Wang, ``{ELM}-based frame
  synchronization in nonlinear distortion scenario using superimposed
  training,'' \emph{IEEE Access}, vol.~9, pp. 53\,530--53\,539, Apr. 2021.

\bibitem{Impef_Ref1}
X.~Zhang, M.~Matthaiou, M.~Coldrey, and E.~Bj$\ddot{\mathrm{o}}$rnson, ``Impact
  of residual transmit {RF} impairments on training-based {MIMO} systems,''
  \emph{IEEE Trans. Commun.}, vol.~63, no.~8, pp. 2899--2911, Aug. 2015.

\bibitem{Impef_Ref2}
X.~Xia, D.~Zhang, K.~Xu, W.~Ma, and Y.~Xu, ``Hardware impairments aware
  transceiver for full-duplex massive {MIMO} relaying,'' \emph{IEEE Trans.
  Signal Process.}, vol.~63, no.~24, pp. 6565--6580, Dec. 2015.

\bibitem{Ref722_5}
Y.~Wang, M.~Liu, J.~Yang, and G.~Gui, ``Data-driven deep learning for automatic
  modulation recognition in cognitive radios,'' \emph{IEEE Trans. Veh.
  Technol.}, vol.~68, no.~4, pp. 4074--4077, Feb. 2019.

\bibitem{Ref5_15}
H.~Zhang, Y.~Li, and Y.~Yuan, ``Practical considerations on channel estimation
  for up-link {MC-CDMA} systems,'' \emph{IEEE Trans. Wireless Commun.}, vol.~7,
  no.~11, pp. 4384--4392, Nov. 2008.

\bibitem{RRef1111}
F.~Alberge, ``Deep learning constellation design for the {AWGN} channel with
  additive radar interference,'' \emph{IEEE Trans. Commun.}, vol.~67, no.~2,
  pp. 1413--1423, Feb. 2019.

\bibitem{Ref2424}
X.~Gao, S.~Jin, C.~Wen, and G.~Y. Li, ``Com{N}et: Combination of deep learning
  and expert knowledge in {OFDM} receivers,'' \emph{IEEE Commun. Lett.},
  vol.~22, no.~12, pp. 2627--2630, Dec. 2018.

\bibitem{a9}
S.~Ioffe and C.~Szegedy, ``Batch normalization: accelerating deep network
  training by reducing internal covariate shift,'' in \emph{Proc. Int. Conf.
  Mach. Learn.}, Mar. 2015, pp. 448--456.

\bibitem{a10}
H.~Huang, W.~Xia, J.~Xiong, J.~Yang, G.~Zheng, and X.~Zhu, ``Unsupervised
  learning-based fast beamforming design for downlink {MIMO},'' \emph{IEEE
  Access}, vol.~7, pp. 7599--7605, Dec. 2019.

\bibitem{a7_add}
L.~Liu, C.~Oestges, J.~Poutanen, K.~Haneda, P.~Vainikainen, F.~Quitin,
  F.~Tufvesson, and P.~D. Doncker, ``The {COST} 2100 {MIMO} channel model,''
  \emph{IEEE Wireless Commun.}, vol.~19, no.~6, pp. 92--99, Dec. 2012.

\bibitem{DeppRef}
E.~Alpaydin, \emph{Neural Networks and Deep Learning}, 2016, pp. 85--109.

\bibitem{a12}
D.~Arpit and Y.~Bengio, ``The benefits of over-parameterization at
  initialization in deep {ReLU} networks,'' 2019, \textit{arXiv:1901.03611.}
  [Online]. Available: https://arxiv.org/abs/1901.03611.

\bibitem{a13}
D.~Pedamonti, ``Comparison of non-linear activation functions for deep neural
  networks on {MNIST} classification task,'' 2018, \textit{arXiv:1804.02763.}
  [Online]. Available: https://arxiv.org/abs/ 1804.02763.

\bibitem{IEERef111}
Y.~Yang, F.~Gao, G.~Y. Li, and M.~Jian, ``Deep learning-based downlink channel
  prediction for {FDD} massive {MIMO} system,'' \emph{IEEE Commun. Lett.},
  vol.~23, no.~11, pp. 1994--1998, Aug. 2019.

\bibitem{IV-1}
D.~Kingma and J.~Ba, ``Adam: a method for stochastic optimization,'' 2014,
  \textit{arXiv:1412.6980}. [Online]. Available:
  https://arxiv.org/abs/1412.6980.

\bibitem{IV-2}
V.~Raj and S.~Kalyani, ``Backpropagating through the air: Deep learning at
  physical layer without channel models,'' \emph{IEEE Commun. Lett.}, vol.~22,
  no.~11, pp. 2278--2281, Nov. 2018.

\bibitem{IV-3}
I.~Goodfellow, Y.~Bengio, and A.~Courville, \emph{Deep learning}, 2016.

\bibitem{a14}
A.~Saleh, ``Frequency-independent and frequency-dependent nonlinear models of
  {TWT} amplifiers,'' \emph{IEEE Trans. Commun.}, vol.~29, no.~11, pp.
  1715--1720, Nov. 1981.

\bibitem{IV-5}
M.~ElHassan, M.~Crussiere, J.~Helard, Y.~Nasser, and O.~Bazzi, ``On providing
  the theoretical {EVM} limit for tone reservation {PAPR} reduction
  technique,'' in \emph{Proc. IEEE SPAWC}, Atlanta, GA, USA, May 2020, pp.
  1--5.

\bibitem{Ref803_1}
J.~Tang, J.~Liang, C.~Han, Z.~Li, and H.~Huang, ``Crash injury severity
  analysis using a two-layer stacking framework,'' \emph{Accident Analysis
  ${\rm{\& }}$ Prevention}, vol. 122, pp. 226--238, Jan. 2019.

\bibitem{a5}
A.~K. Gizzini, M.~Chafii, A.~Nimr, and G.~Fettweis, ``Enhancing least square
  channel estimation using deep learning,'' in \emph{Proc. IEEE Veh. Technol.
  Conf.}, Antwerp, Belgium, May 2020, pp. 1--5.

\end{thebibliography}

%
%

\begin{IEEEbiography}[{\includegraphics[width=1in, height=1.25in, clip, keepaspectratio]{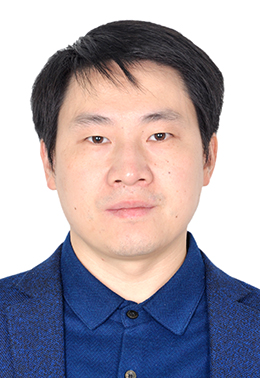}}]{Chaojin Qing} (M'15) received the B.S. degree in communication engineering from Chengdu University of Information Technology, Chengdu, China, in 2001, the M.S. and Ph.D. degrees in communications and information systems from the University of Electronic Science and Technology of China, Chengdu, China, in 2006 and 2011, respectively. From November 2015 to December 2016, he was a Visiting Scholar with Broadband Communication Research Group (BBCR) of the University of Waterloo, Waterloo, ON, Canada.

From 2001 to 2004, he was a teacher with the Communications Engineering Teaching and Research Office, Chengdu University of Information Technology, Chengdu, China. Since 2011, he has been an Assistant Professor with the School of Electrical Engineering and Electronic Information, Xihua University, Chengdu, China. He is the author of more than 50 papers and more than 20 chinese inventions. His research interests include detection and estimation, massive MIMO systems, and deep learning in physical layer of wireless communications.
\end{IEEEbiography}

\begin{IEEEbiography}[{\includegraphics[width=1in, height=1.25in, clip, keepaspectratio]{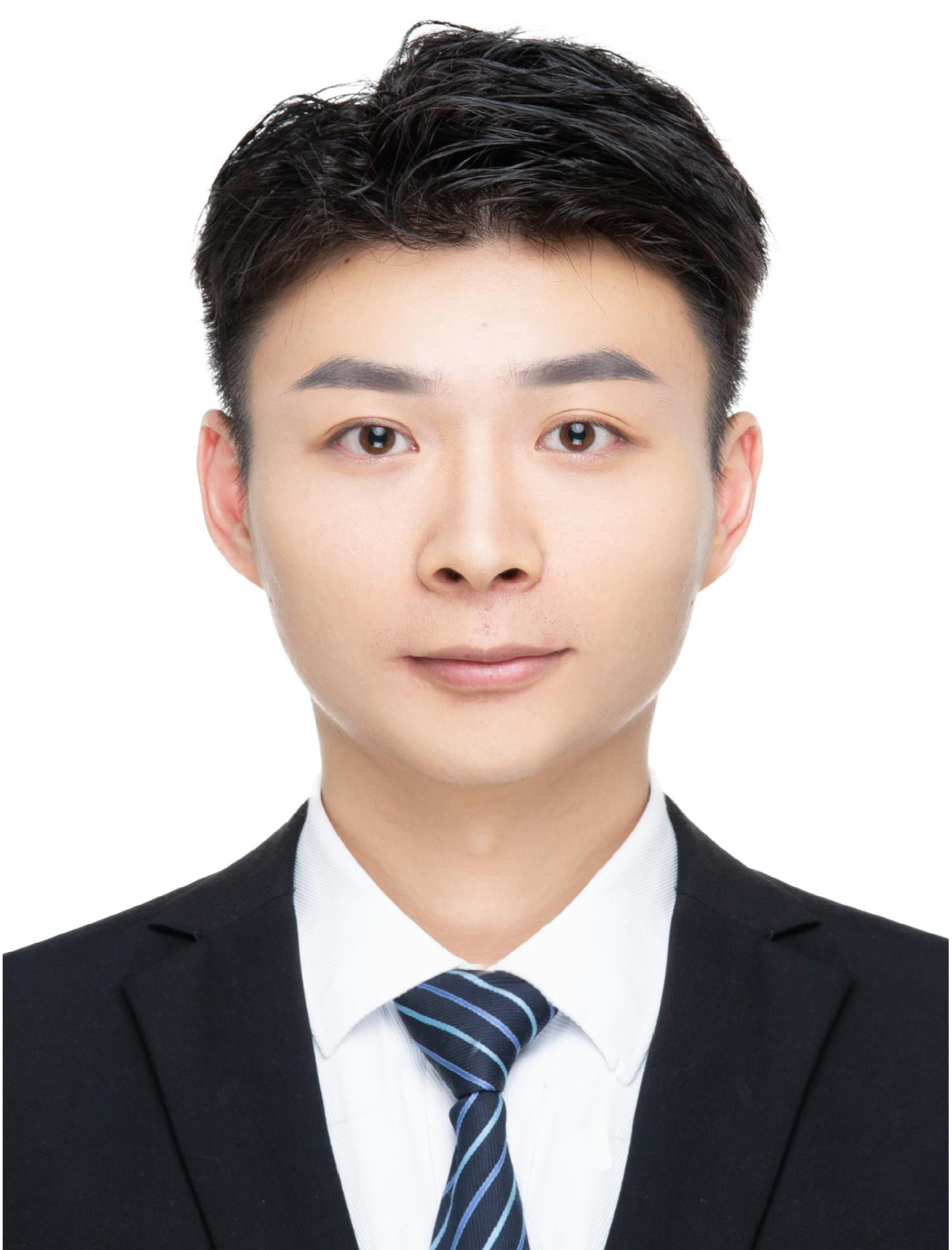}}]{Lei Dong}
received the B. S. degree from the School of Electrical Engineering and Electronic Information, Xihua University, Chengdu, China, in 2019, where he is currently pursuing the M. S. degree under the supervision of Prof. Qing. His research interests include superimposed training based signal detection and channel estimation, and deep learning in physical layer of wireless communications.
\end{IEEEbiography}

\begin{IEEEbiography}[{\includegraphics[width=1in, height=1.25in, clip, keepaspectratio]{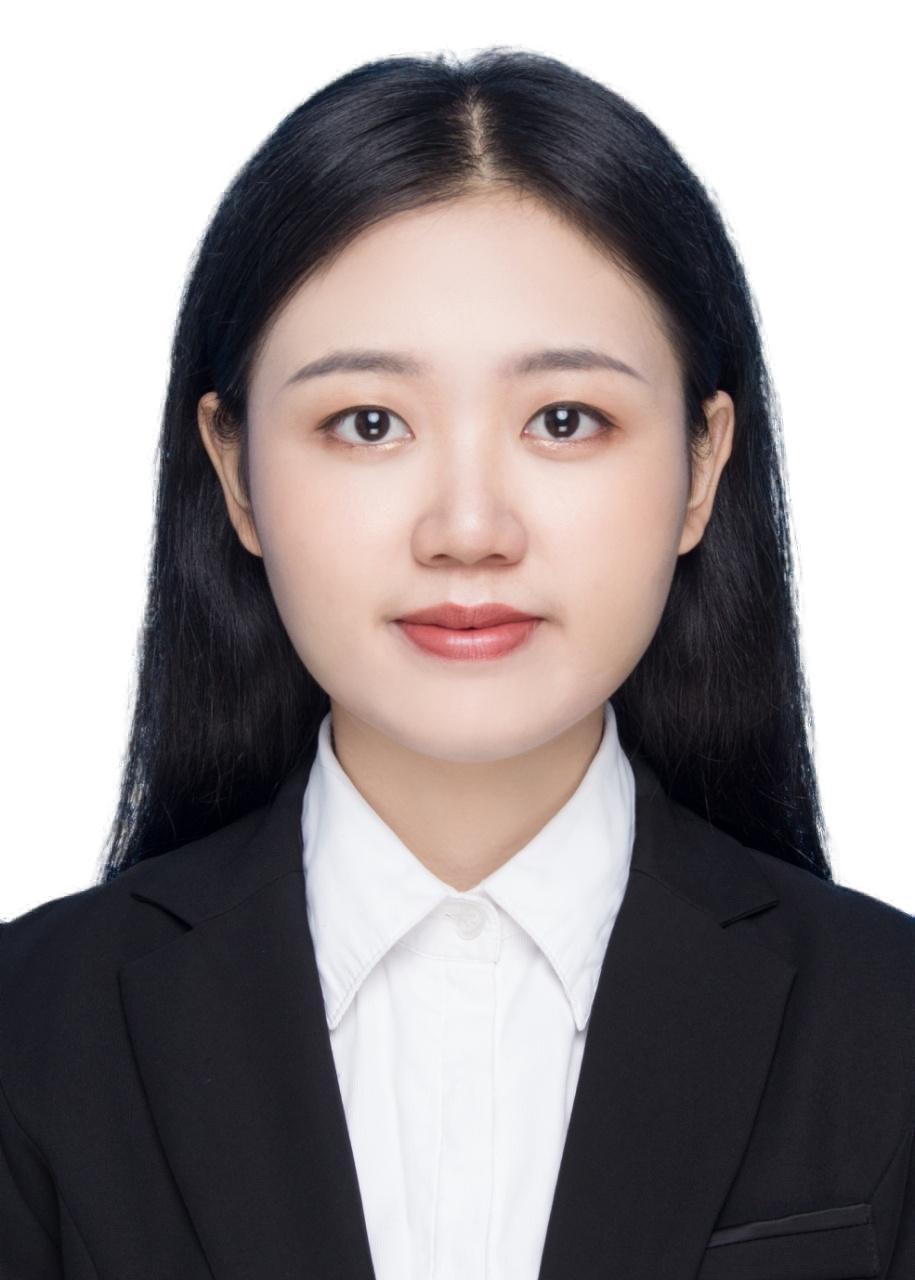}}]{Li Wang} received the B. S. degree from the School of Electrical Engineering and Electrical Information, Xihua University, Chengdu, China, in 2020, where she is currently pursuing the M. S. degree under the supervision of Prof. Qing. Her research interests include channel estimation, signal detection, and deep learning in physical layer of wireless communications.
\end{IEEEbiography}

\begin{IEEEbiography}[{\includegraphics[width=1in, height=1.25in, clip, keepaspectratio]{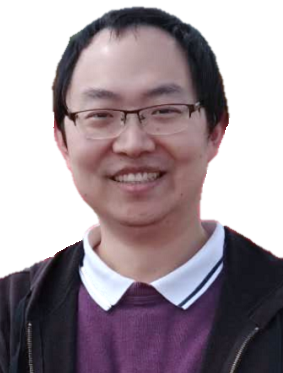}}]{Jiafan Wang} (S'15) received his B.S. degree and M.S. degree in Electrical Engineering from University of Electronic Science and Technology of China in 2006 and 2009, respectively. He accomplished the Ph.D. degree in Computer Engineering at Texas A$\&$M University, College Station, TX, USA in 2017.

His major field of study is smart integrated circuit design, which includes multi-dimensional non-deterministic gate implementation with systematic optimization framework, self-training Analog/Digital Mixed-System for circuit feature calibration, and configurable locking mechanism against Analog IP piracy. He is currently working in Synopsys Inc. to develop the world's leading silicon chip design software in electronic design automation (EDA) industry.
\end{IEEEbiography}

\begin{IEEEbiography}[{\includegraphics[width=1in, height=1.25in, clip, keepaspectratio]{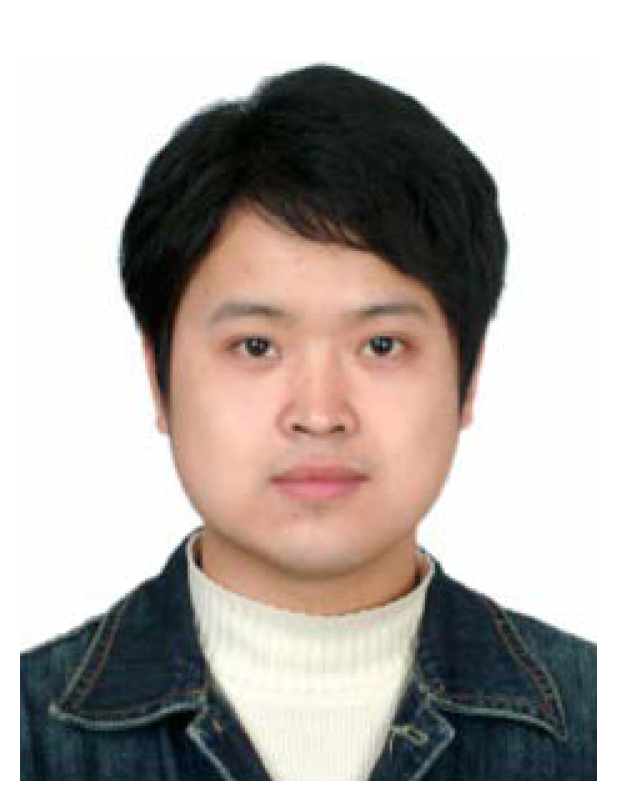}}]{Chuan Huang} (S'09--M'13) received the B.S. degree in math and the M.S. degree in communications engineering from the University of Electronic Science and Technology of China(UESTC), Chengdu, and the Ph.D. degree in electrical engineering from Texas A$\&$M University, College Station, TX, USA, in 2012. From 2012 to 2013, he was with Arizona State University, Tempe, AZ, USA, as a Post-doctoral Research Fellow, and then promoted to Assistant Research Professor from 2013 to 2014. He was also a Visiting Scholar with the National University of Singapore and a Research Associate with Princeton University. From January 2015 to February 2020, he was a professor at UESTC, Chengdu, China.

He is currently an associate professor with the Chinese University of Hong Kong, Shenzhen, China. He served as a Symposium Chair of IEEE GLOBECOM 2019 and IEEE ICCC 2019 and 2020. He is now serving as an Editor of IEEE Transactions on Wireless Communications, IEEE Access, and IEEE Wireless Communications Letters. His current research interests include wireless communications and signal processing.
\end{IEEEbiography}




\end{document}